\documentclass[preprint,3p,12pt]{elsarticle}
\usepackage{lineno,hyperref}

\usepackage{amsmath}
\biboptions{sort&compress}
\usepackage{verbatim} 
\usepackage{epsfig}
\usepackage{amssymb} 
\usepackage{mathtools}
\usepackage{lineno}
\usepackage{tikz}        
\usepackage{etoolbox}
\usepackage{ulem}         
\usepackage{textcomp}
\usepackage{nicematrix}
\usepackage{silence}
\usepackage{hyperref}
\usepackage{float}
\usepackage{subcaption}
\usepackage{graphicx}
\usepackage{placeins}

\newcommand{\Rho}{\mathrm{P}}
\journal{Physica A}
\begin{document}

\newcommand*{\hwplotB}{\raisebox{3pt}{\tikz{\draw[red,dashed,line 
width=3.2pt](0,0) -- 
(5mm,0);}}}

\newrobustcmd*{\mydiamond}[1]{\tikz{\filldraw[black,fill=#1] (0.05,0) -- 
(0.2cm,0.2cm) -- (0.35cm,0) -- (0.2cm,-0.2cm) -- (0.05,0);}}

\newrobustcmd*{\mytriangleright}[1]{\tikz{\filldraw[black,fill=#1] (0,0.2cm) -- 
(0.3cm,0) -- (0,-0.2cm) -- (0,0.2cm);}}

\newrobustcmd*{\mytriangleup}[1]{\tikz{\filldraw[black,fill=#1] (0,0.3cm) 
-- (0.2cm,0) -- (-0.2cm,0) -- (0,0.3cm);}}

\newrobustcmd*{\mytriangleleft}[1]{\tikz{\filldraw[black,fill=#1] (0,0.2cm) -- 
(-0.3cm,0) -- (0,-0.2cm) -- (0,0.2cm);}}
\definecolor{Blue}{cmyk}{1.,1.,0,0} 

\begin{frontmatter}

\title{ Testing lockdown measures in epidemic outbreaks through mean-field models considering the social structure} 

\address[IB]{Instituto Balseiro, Universidad Nacional de Cuyo}
\address[CONICET]{Consejo Nacional de Investigaciones Científicas y Técnicas (CONICET), Bariloche, Argentina}
\address[CNEA]{Gerencia de Física, Centro Atómico Bariloche, Comisión Nacional de Energía Atómica }

\author[IB,CONICET]{E.A.~Rozan}
\author[CONICET,CNEA]{S.~Bouzat}
\author[IB,CONICET,CNEA]{M.N.~Kuperman}

\begin{abstract}
Lately, concepts such as lockdown, quarantine, and social distancing have become very relevant since they have been associated with essential measures in the prevention and mitigation of COVID-19. While some conclusions about the effectiveness of these measures could be drawn from field observations, many mathematical models aimed to provide some clues. However, the reliability of these models is questioned, especially if the social structure is not included in them. In this work, we propose a mesoscopic model that allows the evaluation of the effect of measures such as social distancing and lockdown when the social topology is taken into account. The model is able to predict successive waves of infections without the need to account for reinfections, and it can qualitatively reproduce the wave patterns observed across many countries during the COVID-19 pandemic. Subsequent waves can have a higher peak of infections, if the restrictiveness of the lockdown is above a certain threshold.The model is flexible and can implement various social distancing strategies by adjusting the restrictiveness and the duration of lockdown measures or specifying whether they occur once or repeatedly. It also includes the option to consider essential workers that do not isolate during a lockdown.
\end{abstract}


\end{frontmatter}


\setlength{\parskip}{12pt}
\section{Introduction} \label{section:intro}

During the siege of the COVID-19 pandemic, the concepts of social distancing and quarantine began to transcend the purely clinical and academic fields and permeated the popular imaginary in most countries. Not only was their effectiveness questioned, but they were also associated with conspiracy theories about the desire of the governments to subject the population to repressive state control. The history of quarantine teaches us that such measures were never well received by the masses, which is why a careful evaluation of their effectiveness is necessary.

While the practice of isolating infected people is already mentioned in ancient writings such as the Old Testament, the concept of quarantine, as we know it today, dates back to the 14th century. At the time, its implementation was an effort to protect coastal cities from the fast-spreading plague. Initial outbreaks of plague were immediately associated with the arrival of ships from infected ports, therefore some cities required them to remain at anchor for 40 days before disembarking. One of the first documented measures of this kind was the legislation of the ancient city of Ragusa (Dubrovnik) that required the mandatory isolation of all incoming ships and caravans for 30 days to detect infections \cite{gensi04}.

The current name quarantine derives from the Italian word quarantino, a period of 40 days that was adopted some years later. The number 40 may have been chosen because the number had great symbolic and religious significance to medieval Christians.

In addition to the quarantine, other measures were adopted. Those who were able to isolate themselves from the rest of the population dodged the plague by retirement to isolated villas or palaces. These practices are documented in G. Bocaccio's The Decameron \cite{boca} and in Isaac Newton's retreat to his family farm at Woolsthorpe Manor, where he developed calculus and the theory of gravitation \cite{west83}.

This primary quarantine has evolved into a more sophisticated scheme that includes border controls, contact tracing, and surveillance. While the effectiveness of such measures is unquestionable, these practices have always generated controversy and public debate because they conflict with individual rights and are sometimes abused for political and economic reasons. \cite{togno13}.

While most statistics and field data show that these measures were indeed extremely useful, many researchers sought to support their claims with mathematical models of various kinds. Some models focus on macroscopic perspectives and describe the phenomenon using systems of differential equations \cite{zhang,fuad,liu,giordano,mwal}, while others propose agent-based and network-based models that focus on individuals \cite{svoboda2022,cornes2022,kara}. Each of these treatments has its pros and cons. While a mean-field model allows for the analytic treatment of the problem, it screens the important effect of the social structure. This aspect has proven extremely relevant as shown in Refs.~\cite{moreno2002,dynamical_patterns,R0_heterogeneo,kuperman,zanette,bohme}. Conversely, agent based models and network models require extensive statistics to avoid only showing particular cases. 

In this paper, we propose a conciliatory approach, which takes the most relevant aspects of both proposals. To do this, we propose a model inspired by the one introduced by the authors of~\cite{moreno2002} and further analyzed in~\cite{velocity_spread,dynamical_patterns,zhu}, that is in principle associated with a mean-field approach but incorporates information on the degree distribution of the network that represents the social topology.

Although some concessions must indeed be made along the way, in this work we analyze the results that can be obtained by modeling an epidemiological process according to this procedure. In particular, we are interested in studying the effects that a measure such as social isolation may have, which precisely affects the degree distribution of the social network, restricting the number of contacts that each individual can have.

\section{The model} \label{section:models}
\subsection{The \textit K-SEIR model}\label{section:KSEIR}
The starting point for the present work is the classical SEIR model. In the basic SEIR compartmental mean-field model, the individuals are considered to be in one of four possible states (compartments): S (susceptible to be infected), E (exposed to the infection but not yet infectious ), I (infected, can spread the infection) and R (removed individuals, after either recovering from the disease or dying). The evolution of the infection is described by the following equations:
\begin{equation}
\begin{NiceMatrix}[l]
\displaystyle\frac{\text d S}{\text d t} =&
\displaystyle -r\ SI\\[0.5cm]
\displaystyle \frac{\text d E}{\text d t} =&
\displaystyle r\ SI - \frac{E}{T_\text{inc}}\\[0.5cm]
\displaystyle \frac{\text d I}{\text d t} =&
\displaystyle  \frac{E}{T_\text{inc}} - \frac{I}{T_\text{inf}}\\[0.5cm]
\displaystyle\frac{\text d R}{\text d t} =&
\displaystyle  \frac{I}{T_\text{inf}}\end{NiceMatrix}
\label{SEIR}
\end{equation}
\noindent where $S+E+I+R=1$, as each variable, represents the fraction of the individuals in the respective compartment. $r$ is the mean effective infection rate, $T_\text{inc}$ is the incubation time and $T_\text{inf}$ is the infection time. Our goal is to generalize this model to include a certain degree of detail about the social topology. Since we will not focus on the microscopic level, that is, without describing individual behaviors neither in the epidemiological nor in the social context, we call this a mesoscopic model.

The \textit {K}-SEIR model here presented follows a degree based mean-field approach. In this model, the population is represented as a complex network. The degree distribution $\left(\pi(k),\ k\in\{1,2,...,K\}\right)$ of the individuals (\textit{i.e.}, the fraction of individuals that have $k$ daily contacts) plays a key role. Each of the epidemiological compartments introduced above is divided into ${K}$ sub-compartments, where ${K}$ is the maximum degree attained by any individual. 

For example, the S compartment is divided into S$_1$,S$_2$,...,S$_{{K}}$, where $S_k$ is the fraction of individuals with degree $k$ that are susceptible. Analogously, $E_k$, $I_k$, and $R_k$ are the fractions of individuals with degree $k$ in the respective epidemiological state. This implies $S_k+E_k+I_k+R_k=1\:\forall k$. 

The total fraction of individuals in the $\rho$ compartment (where $\rho$ stands for any of S, E, I or R), is computed as follows:
\begin{equation}
\rho_\text{tot}=\sum_{k=1}^{{K}}\rho_k \pi(k)
 \label{rho_tot}
\end{equation}

Based on the model proposed in \cite{moreno2002} with a slight modification presented by the authors in \cite{velocity_spread}, the equations of the $4{K}$ sub-compartments of the $K$-SEIR model are

\begin{equation}
\begin{NiceMatrix}[l]
\displaystyle\frac{\text d S_k}{\text d t} =&
\displaystyle -r\; \frac{kS_k}{\langle k\rangle}\sum_{k'=1}^K I_{k'}(k'-1)\pi(k')\\[0.5cm]
\displaystyle \frac{\text d E_k}{\text d t} =&
\displaystyle r\; \frac{kS_k}{\langle k\rangle}\sum_{k'=1}^K I_{k'}(k'-1)\pi(k') - \frac{E_k}{T_\text{inc}}\\[0.5cm]
\displaystyle \frac{\text d I_k}{\text d t} =&
\displaystyle  \frac{E_k}{T_\text{inc}}-\frac{I_k}{T_\text{inf}}\\[0.5cm]
\displaystyle\frac{\text d R_k}{\text d t} =&
\displaystyle  \frac{I_k}{T_\text{inf}}\end{NiceMatrix}
\label{KSEIR}
\end{equation}

\subsection{Modeling Social Distancing}

During the pandemic, social distancing was implemented by urging individuals to reduce contact with their social environment. In the context of the ${K}$-SEIR model introduced above, this preventive measure can be represented by changing the degree distribution of the population, so that the individuals decrease their number of daily contacts. This would be reflected in the fact that the proportion of individuals with lower degrees would grow at the expense of the decrease of those corresponding to higher degrees. It is worth noting that we consider this transition to be instantaneous and not gradual. 

As in Ref.~\cite{svoboda2022}, we will consider that the lockdown should start when the total fraction of infected individuals $I_\text{tot}$ is above a given threshold $\tau$. The preventive measure should end when $I_\text{tot}< \tau$ for $d$ consecutive days. In our model, at that point, the degree distribution recovers the original functional form, though the epidemiological profile does not. 

It is important to note that the modeling of any preventive measure can not instantaneously change the epidemiological state of the individual, \textit{i.e.}, the total fraction of individuals in each compartment, $\rho_\text{tot}$, is conserved at the start of the lockdown, even when the degree distribution changes. However, the fractions in each sub-compartment $\rho_k$ are modified. To illustrate that, let us consider that $\pi^\mathrm i(k)$ and $\pi^\mathrm f(k)$ are the degree distributions before and after the lockdown starts, respectively, and $\rho_k^\mathrm i$ and $\rho_k^\mathrm f$ are the fractions in the sub-compartment $\rho_k$ immediately before and after that moment, respectively. Then,
\begin{equation}
\rho_\text{tot}=\sum_k \rho_k^\mathrm i \pi^\mathrm i(k)=\sum_k \rho_k^\mathrm f \pi^\mathrm f(k)
\label{discontinuity}
\end{equation}
As generally $\pi^\mathrm f(k)\neq\pi^\mathrm i(k)$, then $\rho_k^\mathrm f\neq\rho_k^\mathrm i$. This reflects the fact that when individuals change their degrees, the proportion of individuals in each sub-compartment will likely change. For example, when a lockdown starts we expect a higher fraction of individuals with degree $k=2$ ($\pi^\mathrm f(2)>\pi^\mathrm i(2)$), so the new proportion of susceptible individuals with degree $k=2$ will likely change: $S_2^\mathrm i\neq S_2^\mathrm f$. 

To find $\rho_k^\mathrm f$ so that Eq.~(\ref{discontinuity}) holds true, we split the degrees into three groups: 
\begin{itemize}
    \item Group \textbf{I} contains the degrees $j$ whose probability increases: $\pi^\mathrm f(j)>\pi^\mathrm i(j)$.
    \item Group \textbf{II} contains the degrees $k$ whose probability is maintained: $\pi^\mathrm f(k) = \pi^\mathrm i(k)$.
    \item Group \textbf{III} contains the degrees $l$ whose probability decreases: $\pi^\mathrm f(l)<\pi^\mathrm i(l)$.
    \end{itemize}
If $f_l=\pi^\mathrm i(l)-\pi^\mathrm f(l)$ is the fraction of individuals that leave the compartments with degree $l$ in group \textbf{III}; and $g_j=\pi^\mathrm i(l)-\pi^\mathrm f(l)$ is the fraction of individuals that are incorporated to the compartments with degree $j$ in group \textbf{I},then
\begin{equation}
\sum_l f_l=\sum_j g_j\doteq \Phi
\end{equation}
We consider that in group \textbf{III} the fraction of individuals in each sub-compartment does not change, because individuals that proceed to isolate themselves do so maintaining the proportions $\rho_k$, \textit{i.e.}, the fraction of individuals that leave from degree $l$ and are in the $\rho$ state is $f_l\rho_l^\mathrm i$, and thus it follows that $\rho_l^\mathrm f=\rho_l^\mathrm i$. The same is true for the degrees contained in group \textbf{II}. However, for the degrees in group \textbf{I}, the fractions change because individuals arrive from different degrees $l$ with different proportions $\rho_l^\mathrm i$. 

As a first approximation, we consider the density of individuals that goes from degree $l$ to degree $j$ to be proportional to $f_l$ and $g_j$: 
\begin{equation}
\sigma_{l,j}=\frac{f_l\ g_j}{\Phi}
\end{equation}

Then, calling $\Rho_j^\mathrm i=\rho_j^\mathrm i\pi^\mathrm i(j)$ the initial total fraction of individuals with degree $j$ in compartment $\rho$, and $\Rho_j^\mathrm f=\rho_j^\mathrm f\pi^\mathrm f(j)$ the final one, 
\begin{equation}
\Rho_j^\mathrm f=\Rho_j^\mathrm i+\sum_l\sigma_{l,j}\,\rho^\mathrm i_l
\end{equation}
\noindent and lastly we can find $\rho_j^\mathrm f$:

\begin{equation}
 \rho^\mathrm f_j=\frac{\Rho^\mathrm f_j}{\pi^\mathrm f(j)}
\end{equation}

When a lockdown measure starts or ends, the population is redistributed among the degrees to get the proper degree distribution. In other words, every time the degree distribution changes, the fractions $\rho_j^\mathrm f$ must be calculated by following these steps.

\section{Results}\label{section:results}

In this work, we focused on social topologies modeled according to two particular types of networks: regular networks, and scale-free networks. 

In regular networks, all of the individuals have the same number of contacts, \textit{i.e.} the same degree $k=k_0$. The degree distribution that reflects this behavior is $\pi(k)=\delta_{k,k_0}$. In our model, during a lockdown measure it changes to $\pi(k)=\delta_{k,k_1}$ with $k_1<k_0$. As in Ref.~\cite{svoboda2022}, we will keep $k_0=15$ fixed in this section and vary $k_1$: more restrictive lockdown measures imply lower $k_1$ values. 

In scale-free networks, the degree distribution of the population follows a power-law function: $\pi(k)=\alpha_0k^{-\gamma_0}$. During a lockdown, we assume the distribution to change to $\pi(k)=\alpha_1k^{-\gamma_1}$ with $\gamma_1>\gamma_0$. We will maintain $\gamma_0=2.5$, and vary $\gamma_1$: higher values mean more restrictive social distancing. 

The parameters used for the ${K}$-SEIR model from Section~\ref{section:KSEIR} are ${T_\text{inc}=5.2}$ days and $T_\text{inf}=14-T_\text{inc}=8.8$ days, as these were the initially reported values for the COVID-19 pandemic~\citep{early_transmission_dynamics}. For regular networks $r=0.05$, while for scale-free networks $r=0.2$ and ${K}=200$. The initial conditions are $S_k(0) = 0.99$, $I_k(0) = 0.01, E_k(0)=R_k(0)=0\ \forall k$ in all cases.

For the parameters characterizing the lockdown measures, we analyzed the cases where $d=7$ and $d=70$ as in Ref.~\cite{svoboda2022}. For the threshold of simultaneously infected individuals that triggers a lockdown measure, we used $\tau=0.1$ for regular networks and $\tau=0.05$ for scale-free networks. 

Details on the values chosen for all of the mentioned parameters can be found in the supplementary material accompanying this article, in which we explore the parameter space for each type of network and calculate the respective values of $R_0$.

\begin{figure}[th]\centering
 \begin{subfigure}{.49\textwidth} \centering
 \includegraphics[width=\linewidth]{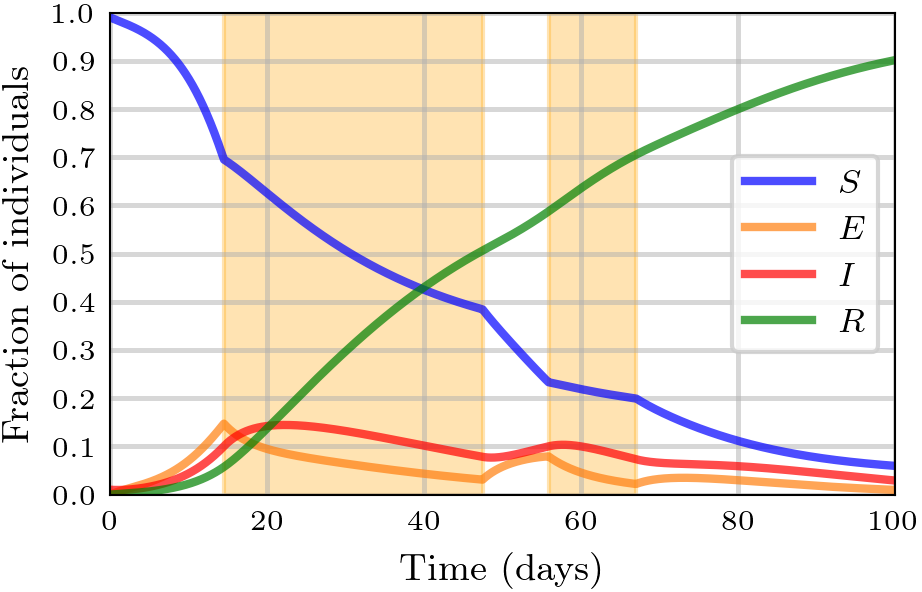}
 \caption{$k_1=4$, successive lockdowns.}
 \end{subfigure}
 \begin{subfigure}{.49\textwidth} \centering
 \includegraphics[width=\linewidth]{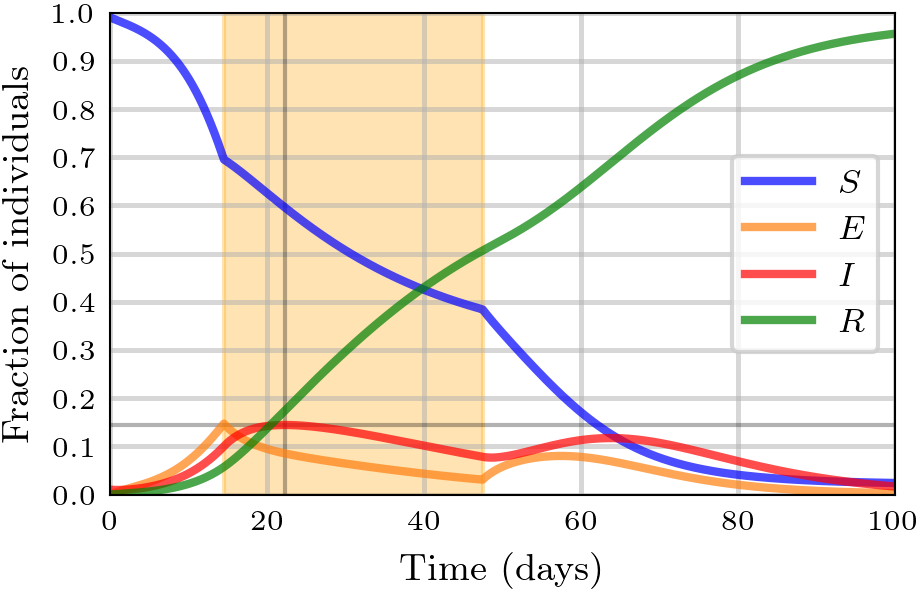}
 \caption{$k_1=4$, only one lockdown.}
 \end{subfigure}
 \begin{subfigure}{.49\textwidth} \centering
 \includegraphics[width=\linewidth]{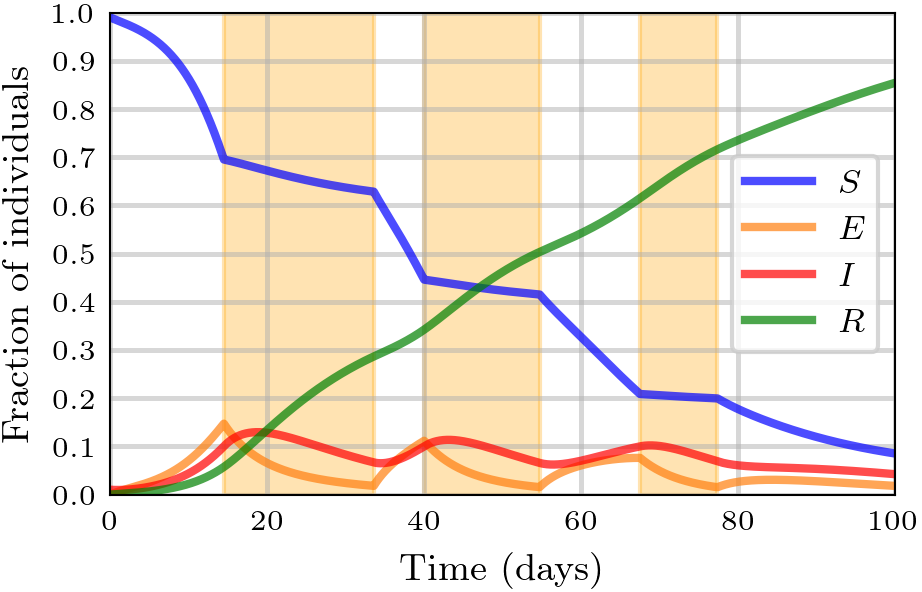}
 \caption{$k_1=2$, successive lockdowns.}
 \end{subfigure}
 \begin{subfigure}{.49\textwidth} \centering
 \includegraphics[width=\linewidth]{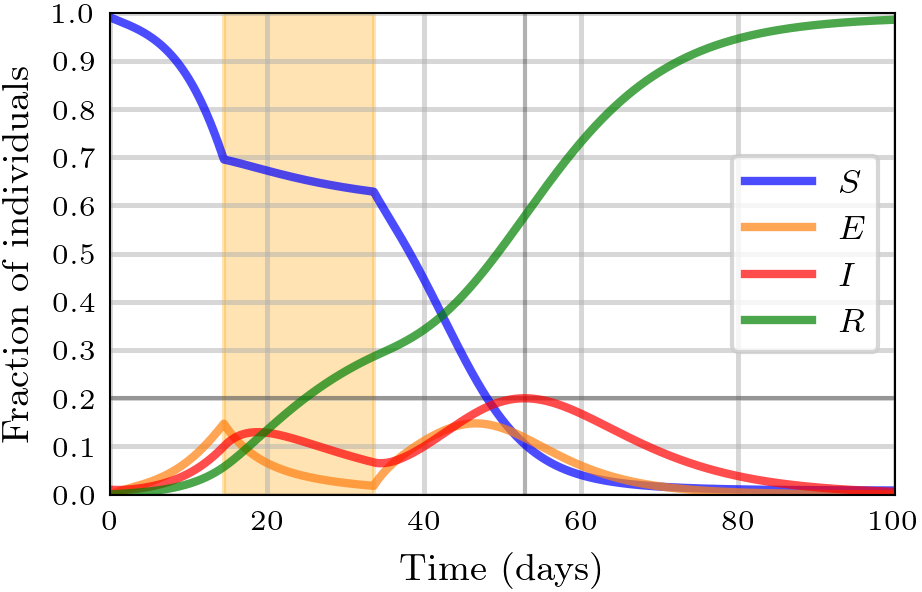}
 \caption{ $k_1=2$, only one lockdown}
 \end{subfigure}
 \caption{Evolution of the total fraction of individuals in each epidemiological compartment in a regular network with $k_0=15$. When ${I>\tau=0.1}$ for the first time, a lockdown starts, and it will last until $I<\tau$ for ${d=7}$ consecutive days. Highlighted in orange are the times for which the lockdown is in place. Depending on the case, the lockdown is allowed to be repeated or not (see text for details). }
 \label{regular_lockdown}
\end{figure}

The evolution of the fraction of individuals in each compartment is shown in Figure~\ref{regular_lockdown}, where the degree distribution corresponds to a regular network. The parameters of the lockdown measure are $d=7\,$days, and $\tau=0.1$. In the top row, $k_1=4$ while in the bottom one $k_1=2$ (more restrictive). In the left column, the lockdown is repeated if $I_\text{tot}>\tau$ after the first wave, and on the contrary, in the right column it can only happen once. 

Figure ~\ref{regular_lockdown} shows that the behavior qualitatively changes if the lockdown measure is allowed to be repeated. We will first focus on the case of the non-repeating lockdown in Section~\ref{section:non_repeating} and will move on to successive lockdowns in Section~\ref{section:succesive}.

\subsection{Non-repeating isolation strategies}\label{section:non_repeating}

\begin{figure}[th]\centering
 \begin{subfigure}{.49\textwidth} \centering
 \includegraphics[width=\linewidth]{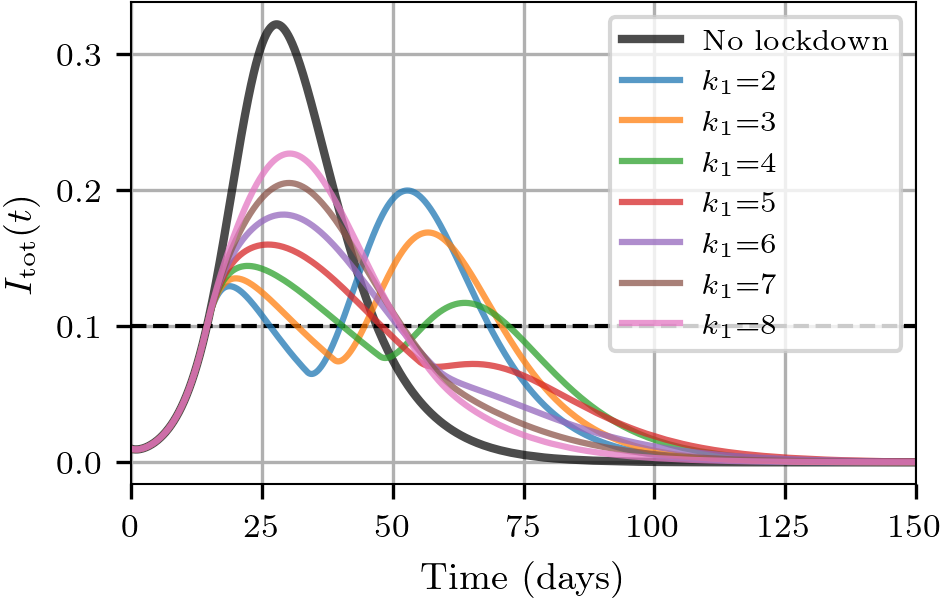}
 \caption{Regular networks}
 \label{infected_norepe_d7_a}
 \end{subfigure} 
 \begin{subfigure}{.49\textwidth} \centering
 \includegraphics[width=\linewidth]{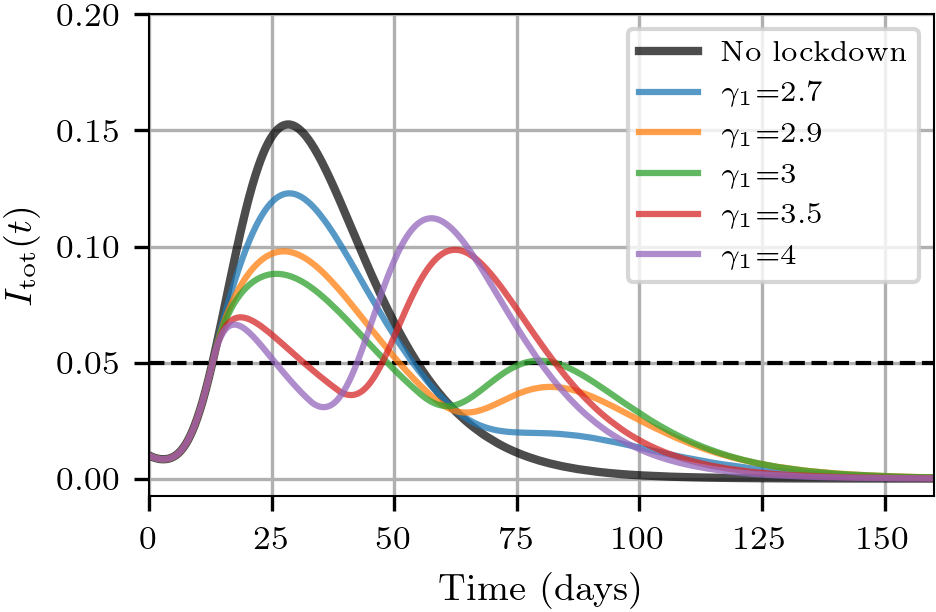}
 \caption{Scale-free networks}
 \label{infected_norepe_d7_b}
 \end{subfigure}
 \caption{$I_\text{tot}(t)$ for different values of $k_1$ and $\gamma_1$. the lockdowns have $d=7$, and it is not repeated even if $I_\text{tot}>\tau$ after the first preventive measure ends. The dashed line represents $I_\text{tot}=\tau$. }
\label{infected_norepe_d7}
\end{figure}

As the most notable changes between the different isolation strategies occur in the evolution of $I_\text{tot}(t)$, in Figure~\ref{infected_norepe_d7} we show some of these curves for different values of $k_1$ and $\gamma_1$ for lockdowns with $d=7$. 

As the confinement gets more restrictive (lower $k_1$, higher $\gamma_1$), the first peak of infections decreases. However, if it gets too restrictive, the number of infections will rapidly decrease, the first peak of infections will come faster (albeit it being at a lower number of infections), and the lockdown will also end faster. Thus, the fraction of susceptible individuals at the end of the lockdown increases for higher restrictiveness, and a second wave of infections will affect them shortly after the preventive measure ends. The height of the second peak will also increase because of this. Furthermore, for $k_1<3.5$ and $\gamma_1>3.2$ the second peak will surpass the first one. This means that it is not always convenient to have more restrictive lockdowns, even when looking at strictly epidemiological variables.

Another notable difference between the results for both types of networks is that in Figure~\ref{infected_norepe_d7_a} there are two waves only if $k_1\leq 5$. With higher $k_1$, the lockdown is more permissive, and more people are infected during its duration, leaving a lower fraction of susceptible individuals at its end, and these are not enough to produce a new wave of infections. Meanwhile, in Figure~\ref{infected_norepe_d7_b}, for lockdown measures with restrictiveness as low as $\gamma_1=2.7$ (remember that $\gamma_0=2.5$) there is a slight increase of infections after the preventive measure ends. Throughout the following sections, this pattern will be repeated and will be discussed in detail in Section~\ref{section:discussion}. 

\begin{figure}[th]\centering
 \begin{subfigure}{.49\textwidth} \centering
 \includegraphics[width=\linewidth]{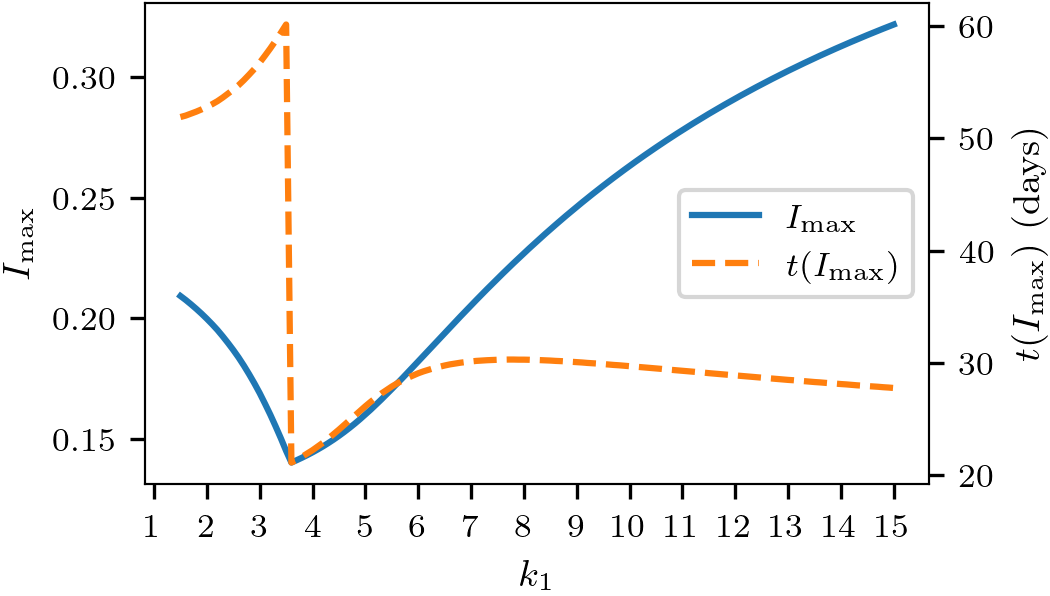}
 \caption{Regular networks}
 \end{subfigure}
 \begin{subfigure}{.49\textwidth} \centering
 \includegraphics[width=\linewidth]{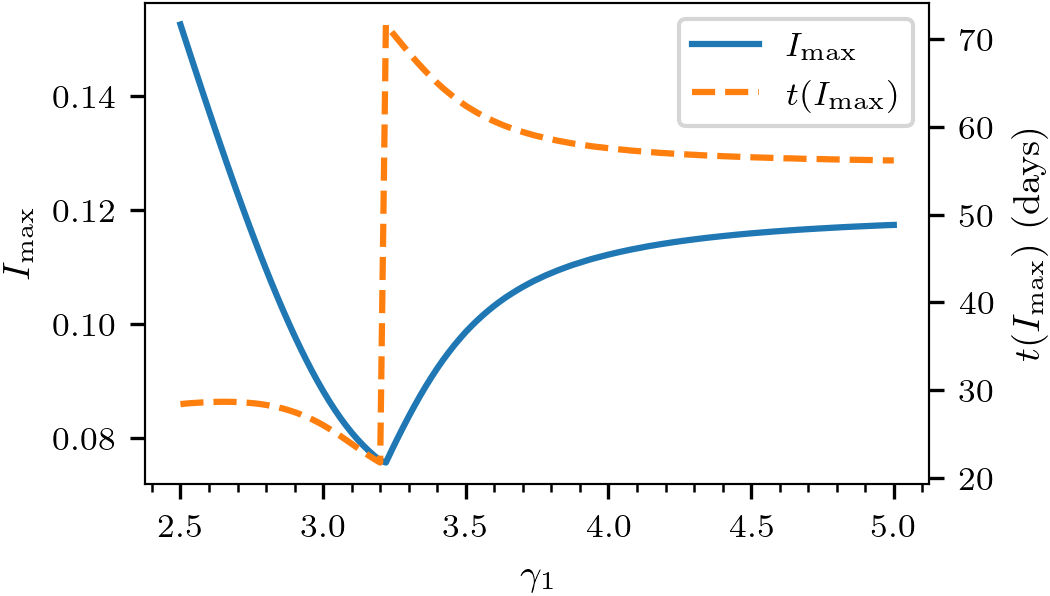}
 \caption{Scale-free networks}
 \end{subfigure}
 \caption{$I_\text{max}$ and $t(I_\text{max})$ as a function of the restrictiveness of the lockdown measures with $d=7$. }
\label{observables_norepe_d7}
\end{figure}

To better visualize how restrictiveness affects the time and height of the highest peak of infections, Figure~\ref{observables_norepe_d7} shows these two quantities as a function of $k_1$ and $\gamma_1$. When the lockdown measures are not overly strict ($k_1\geq3.5$, $\gamma_1\leq3.2$), the first peak is higher and decreases with the level of restriction, and it also comes sooner. After the critical values $k_1=3.5$ and $\gamma_1=3.2$, the second peak is higher, and further increasing the restrictiveness will increase its height.

\begin{figure}[th]\centering
 \begin{subfigure}{.49\textwidth} \centering
 \includegraphics[width=\linewidth]{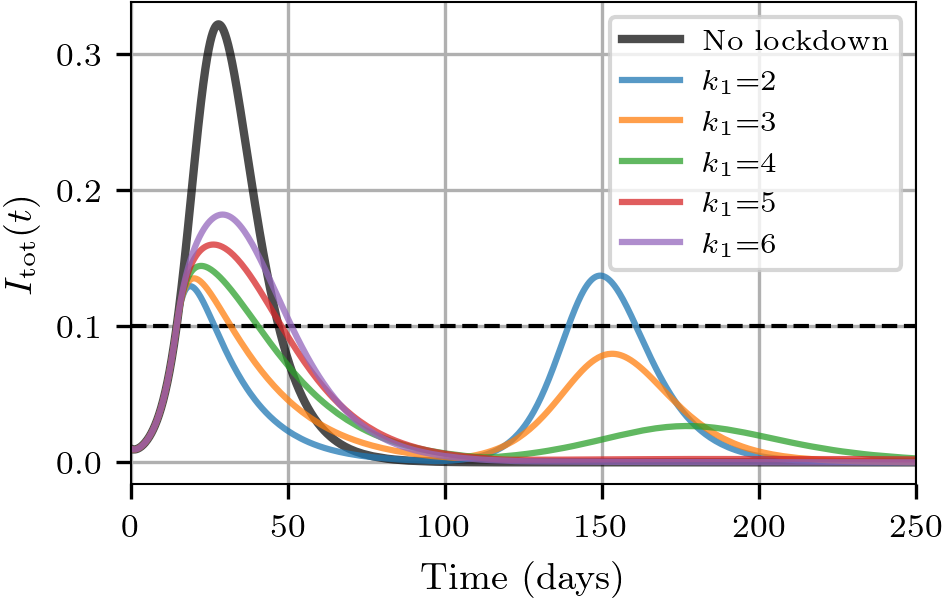}
 \caption{Regular networks}
 \end{subfigure}
 \begin{subfigure}{.49\textwidth} \centering
 \includegraphics[width=\linewidth]{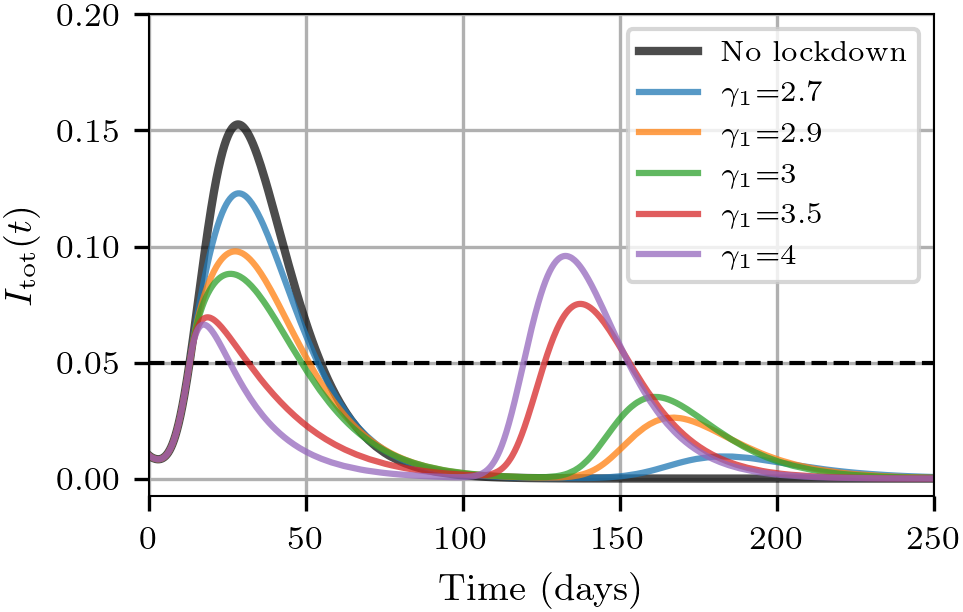}
 \caption{Scale-free networks}
 \end{subfigure}
 \caption{$I_\text{tot}(t)$ for different values of $k_1$ and $\gamma_1$. the lockdowns have $d=70$, and it is not repeated even if $I_\text{tot}>\tau$ after the first preventive measure ends. The dashed line represents $I_\text{tot}=\tau$.}
\label{infected_norepe_d70}
\end{figure}

\begin{figure}[th]\centering
 \begin{subfigure}{.49\textwidth} \centering
 \includegraphics[width=\linewidth]{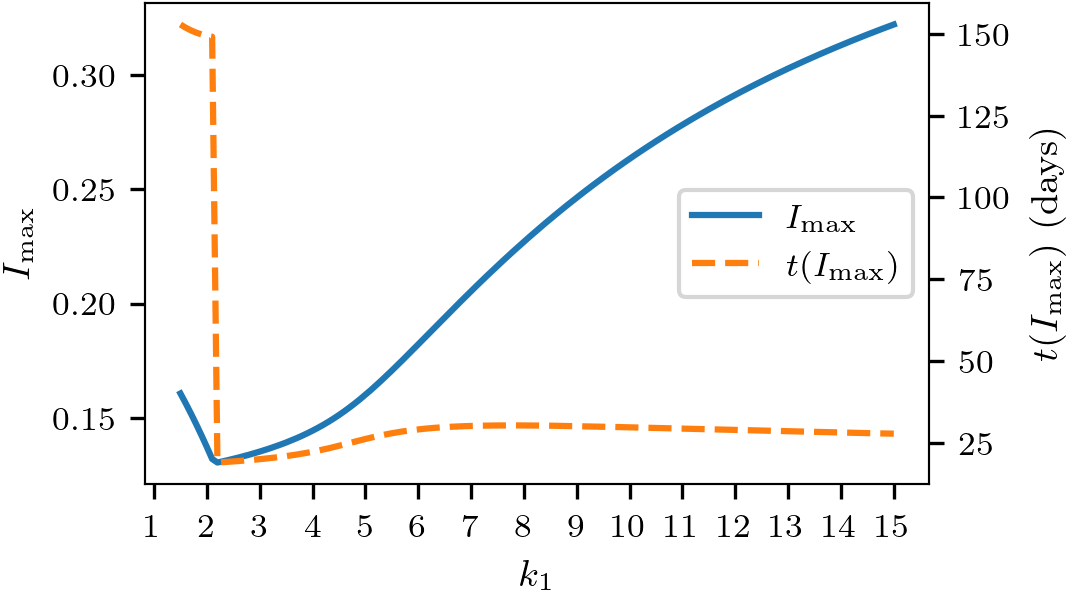}
 \caption{Regular networks}
 \end{subfigure}
 \begin{subfigure}{.49\textwidth} \centering
 \includegraphics[width=\linewidth]{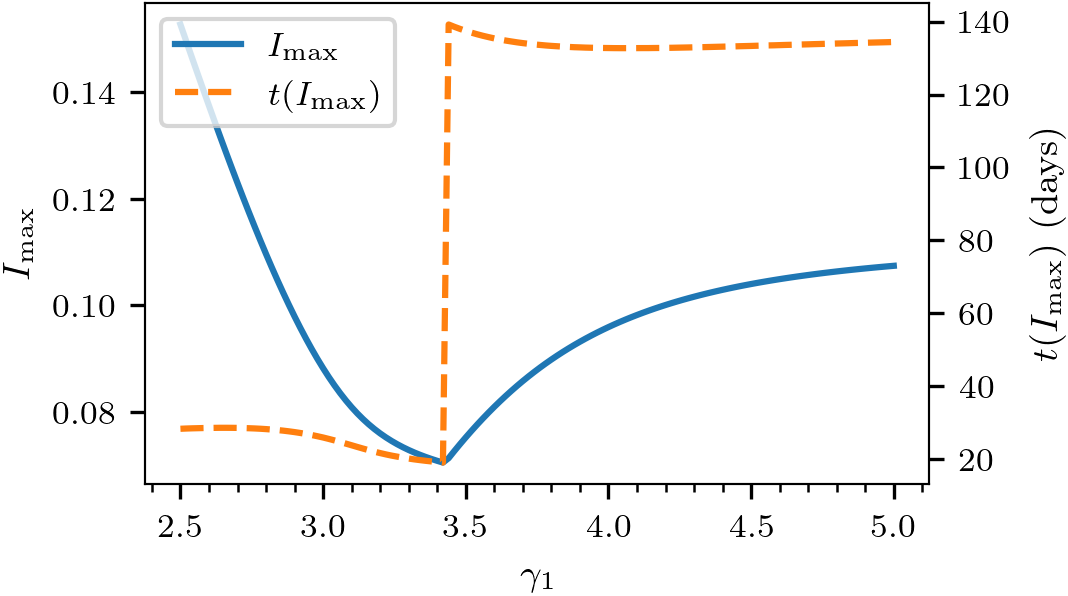}
 \caption{Scale-free networks}
 \label{observables_norepe_d70_b}
 \end{subfigure}
 \caption{$I_\text{max}$ and $t(I_\text{max})$ as a function of the restrictiveness of the lockdown measures with $d=70$.}
\label{observables_norepe_d70}
\end{figure}

Next, we will show the case with $d=70$, which lets each wave of infection be more separated in time and helps us understand their dynamics. Figure~\ref{infected_norepe_d70} shows $I(t)$ for several $k_1$ and $\gamma_1$, while Figure~\ref{observables_norepe_d70} shows $I_\text{max}$ and $t(I_\text{max})$ as a function of the restrictiveness of the lockdown. 

In this case, as the lockdown lasts longer, by the time it ends a considerably higher amount of infections will have happened compared to the previous case with $d=7$, and thus there will be fewer susceptible individuals when it ends. In consequence, the critical value for $k_1$ will be lower and for $\gamma_1$ it will be higher. For regular networks, only with $k_1\leq 2$ the second peak is higher than the first one, while for scale-free networks the critical value is $\gamma_1=3.42$.

Following the pattern commented above, in the case of regular networks a second wave of infections arises only for $k_1\leq4$, while on the other hand for scale-free networks this second wave happens for all values of $\gamma_1$. This difference between both types of networks is seen more clearly in Figure~\ref{infected_norepe_d70} than in Figure~\ref{infected_norepe_d7}, and will be discussed in Section~\ref{section:discussion}.

\subsection{Successive social distancing strategies}\label{section:succesive}

\begin{figure}[th]\centering
 \begin{subfigure}{.49\textwidth} \centering
 \includegraphics[width=\linewidth]{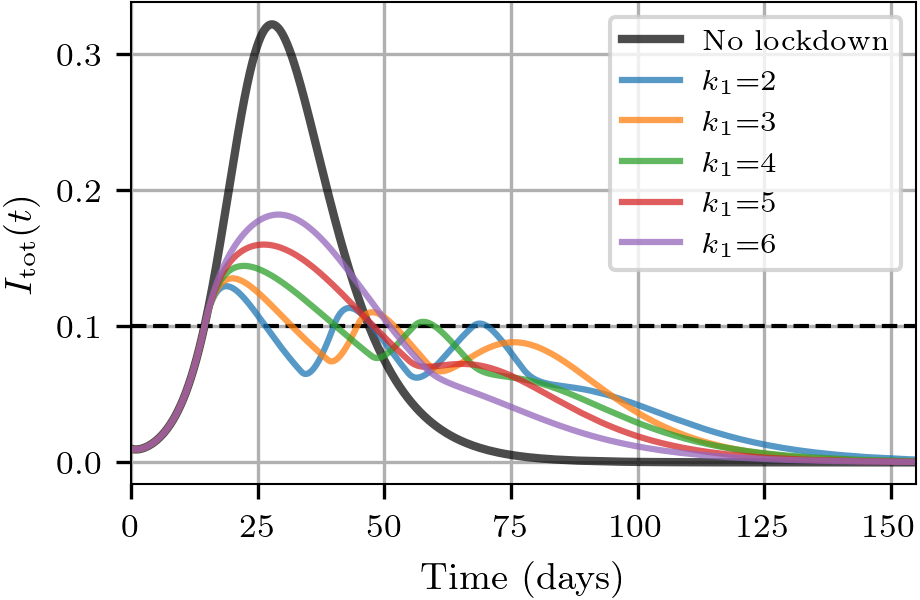}
 \caption{Regular networks}
 \end{subfigure}
 \begin{subfigure}{.49\textwidth} \centering
 \includegraphics[width=\linewidth]{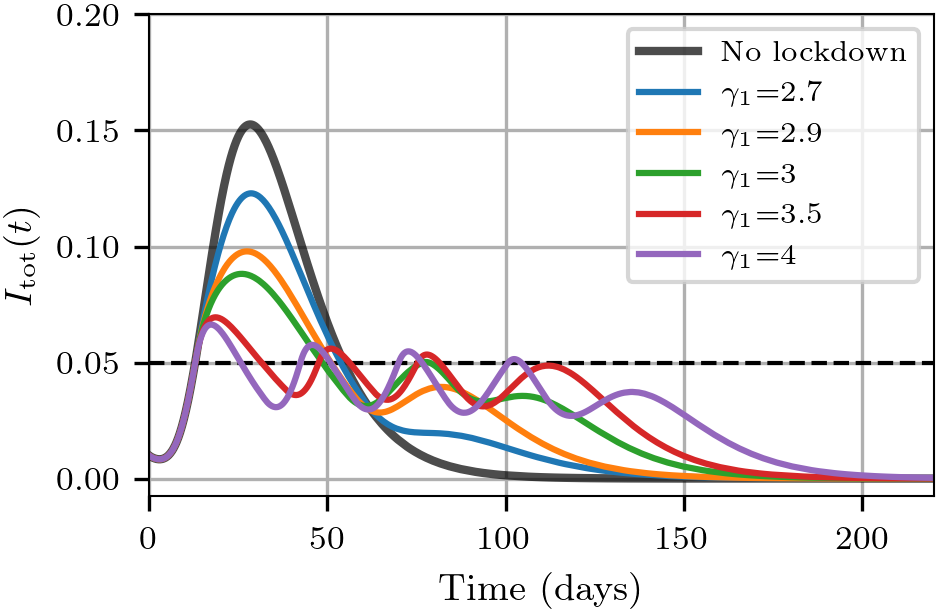}
 \caption{Scale-free networks}
 \end{subfigure}
 \caption{$I_\text{tot}(t)$ for different values of $k_1$ and $\gamma_1$. The successive lockdowns have $d=7$. The dashed line represents $I_\text{tot}=\tau$.}
 \label{infected_repe_d7}
\end{figure}

In this Section, the lockdown measures can be repeated successively if $I_\text{tot}>\tau$ after the first lockdown ends. Figure~\ref{infected_repe_d7} shows the curve of $I_\text{tot}(t)$ for several values of $k_1$ and $\gamma_1$, in both cases for $d=7$. 

The behavior of the first peak is similar to that discussed in the previous section, as it occurs during the first wave. For regular networks, only for $k_1\leq2.5$ there are three separate lockdowns, while for $2.5\leq k_1\leq 4.3$ there are two. For scale-free networks, there are two lockdowns when $3\leq\gamma_1\leq 3.16$, three when $3.16<\gamma_1\leq3.52$, and four when $\gamma_1>3.52$.

\begin{figure}[th]\centering
\begin{subfigure}{.49\textwidth} \centering
\includegraphics[width=\linewidth]{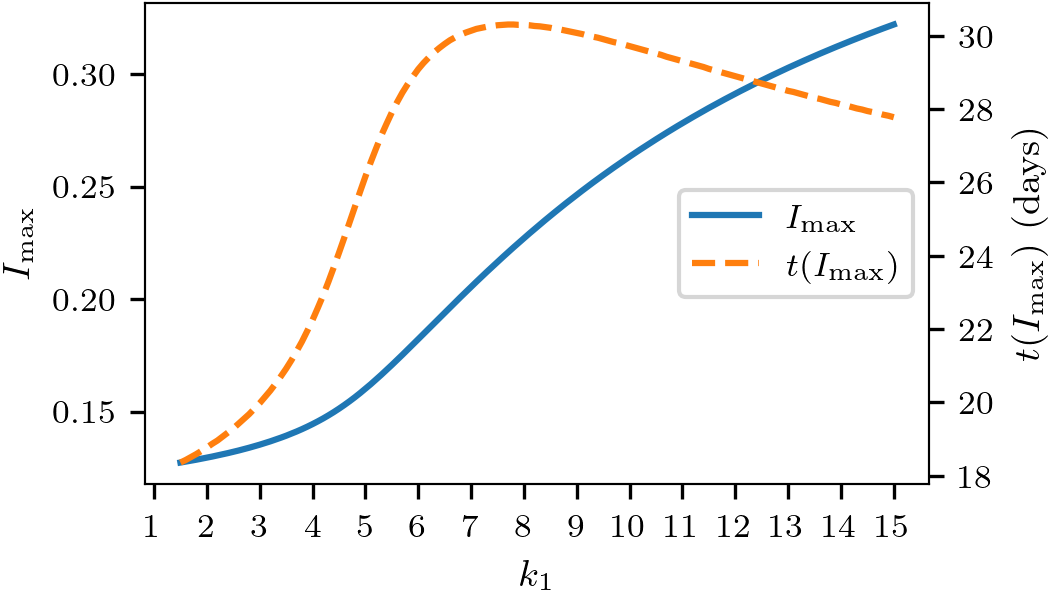}
 \caption{Regular networks}
\label{observables_repe_d7_a}
\end{subfigure}
\begin{subfigure}{.49\textwidth} \centering
\includegraphics[width=\linewidth]{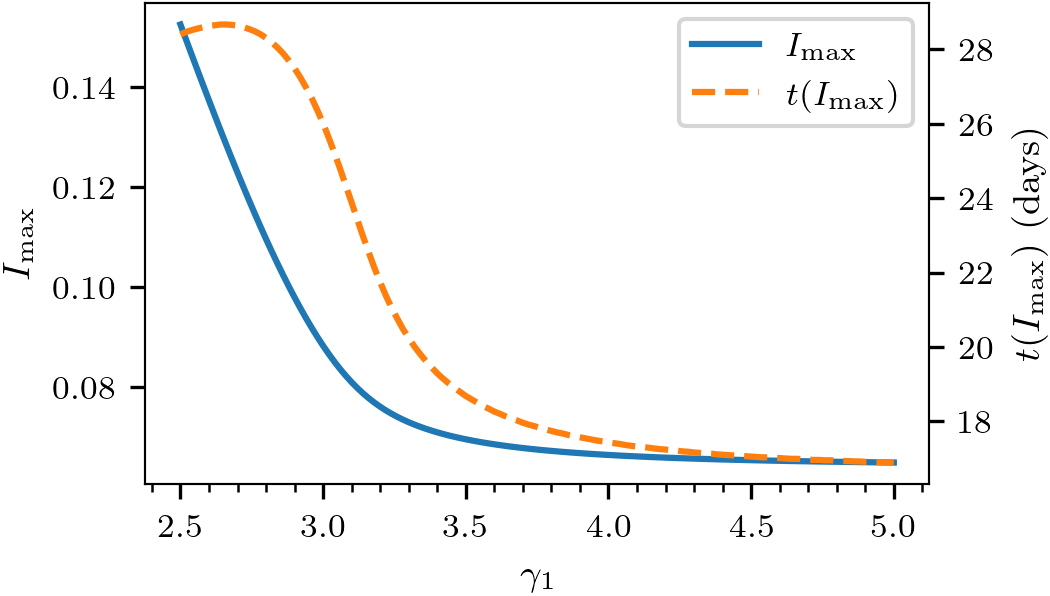}
 \caption{Scale-free networks}
\label{observables_repe_d7_b}
\end{subfigure}
\caption{$I_\text{max}$ and $t(I_\text{max})$ as a function of the restrictiveness of the successive lockdown measures.}
\label{observables_repe_d7}
\end{figure}

Comparing Figure~\ref{infected_norepe_d7} with Figure~\ref{infected_repe_d7}, the main difference is that the first peak is always the highest. This is because every wave that surpasses the threshold $\tau$ will trigger a new lockdown measure, and as each wave begins with a lower fraction of susceptible individuals, their peaks cannot be higher than the first one. As such, $I_\text{max}$ decreases for higher restrictiveness, as shown in Figure~\ref{observables_repe_d7}. On the same note, the behavior of the epidemiological variables presented in Figure~\ref{observables_repe_d7} does not depend on the value of $d$, as the plotted quantities depend only on the behavior of the first wave. 

\begin{figure}[th]\centering
 \begin{subfigure}{.49\textwidth} \centering
 \includegraphics[width=\linewidth]{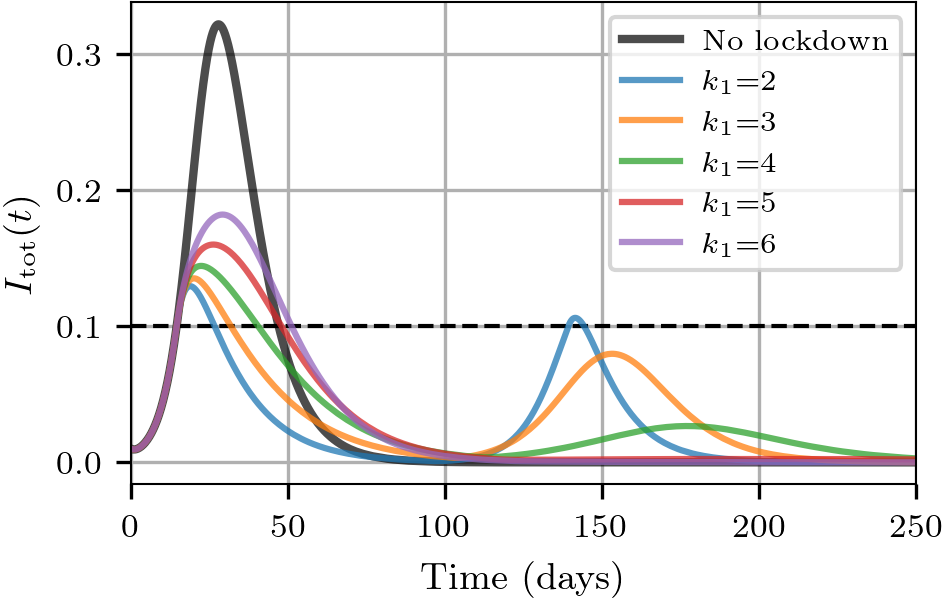}
 \caption{Regular networks}
 \end{subfigure}
 \begin{subfigure}{.49\textwidth} \centering
 \includegraphics[width=\linewidth]{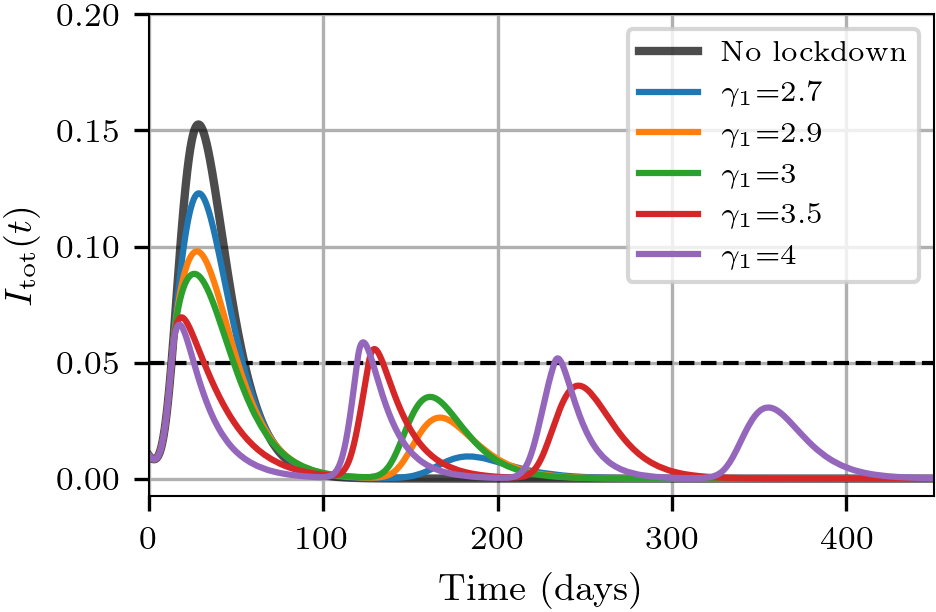}
 \caption{Scale-free networks}
 \end{subfigure}
 \caption{$I_\text{tot}(t)$ for different values of $k_1$ and $\gamma_1$. The successive lockdowns have $d=70$. The dashed line represents $I_\text{tot}=\tau$.}
 \label{infected_repe_d70}
\end{figure}

The differences mentioned above become more apparent when considering a larger value of $d$, as shown in Figure~\ref{infected_repe_d70}. For regular networks for $k_1\leq2.6$ there are two successive lockdowns while for scale-free networks there are two for $3.18\leq \gamma_1\leq3.72$ and three for $\gamma_1>3.72$. On another note, in the case of regular networks, for $k_1\geq 5$, there are no subsequent waves after the lockdown ends. However, in the case of scale-free networks, there is always an additional wave of infections after the last lockdown ends. Specifically, if there is only one lockdown, there are two waves of infections, when there are two lockdowns, a third wave arises, and if there are three lockdowns, a fourth wave takes place. This is in concordance with the ongoing pattern mentioned above and will add to the discussion of Section~\ref{section:discussion}.

\FloatBarrier

\begin{figure}[th]\centering
 \begin{subfigure}{.47\textwidth} \centering
 \includegraphics[width=\linewidth]{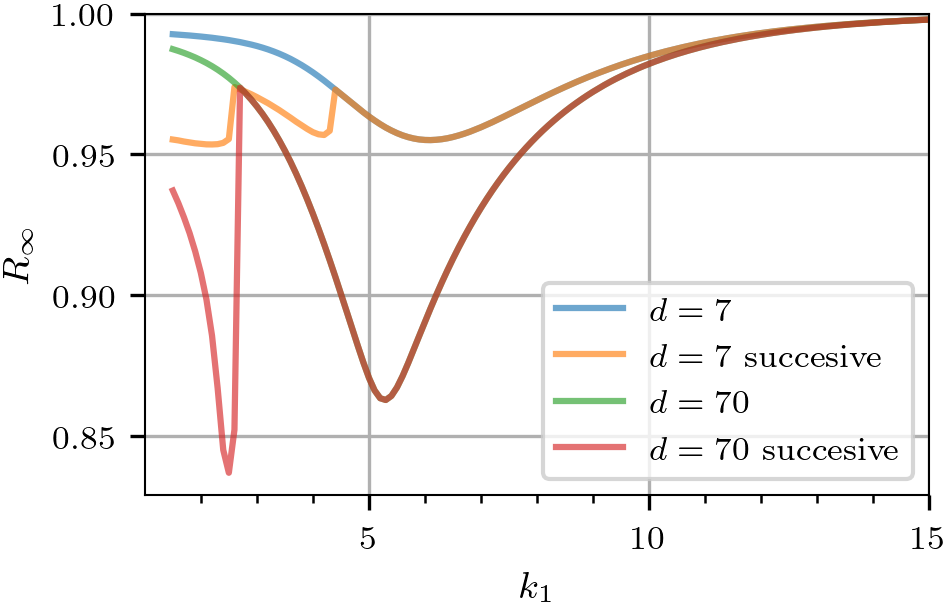}
 \caption{Regular networks}
 \end{subfigure}
 \begin{subfigure}{.52\textwidth} \centering
 \includegraphics[width=\linewidth]{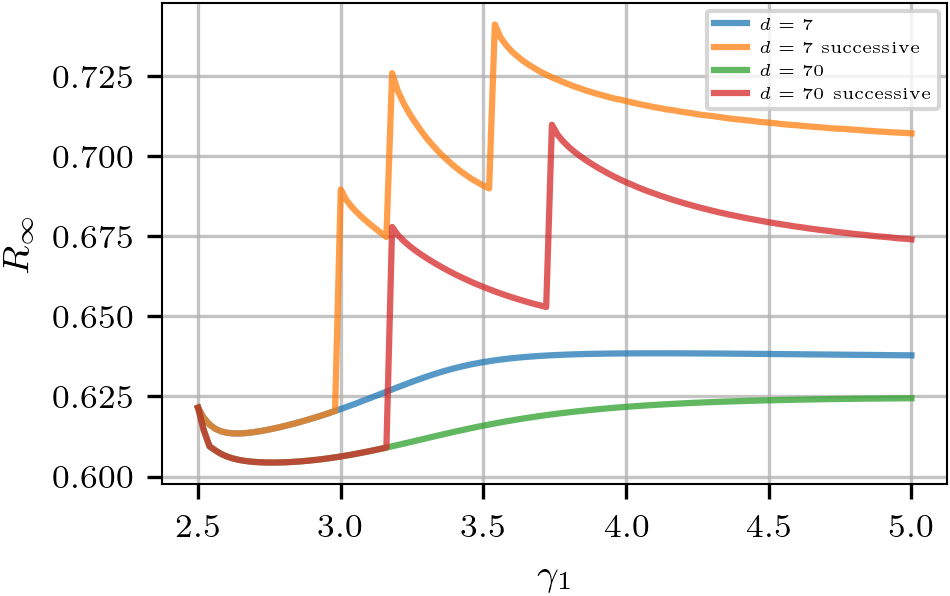}
 \caption{Scale-free networks}
 \label{R_infinity_b}
 \end{subfigure}
 \caption{$R_\infty$ as a function of $k_1$ and $\gamma_1$ in all of the commented scenarios: when the lockdown is repeated, when it is not, and with $d=7$ and $d=70$. }
 \label{R_infinity}
\end{figure}

The most notable difference between regular and scale-free networks, in terms of the epidemiological variables, is the behavior of $R_\infty=R_\text{tot}(t=\infty)$, which quantifies the total number of infections at the end of the epidemic process. It is shown in Figure~\ref{R_infinity} for all of the explored cases: repeating and non-repeating lockdowns, for $d=7$ and $d=70$.

For regular networks, extremely restrictive lockdowns are counterproductive, as discussed above. That is why for $k_1\leq6.1$ with $d=7$ and $k_1\leq 5.3$ with $d=70$, $R_\infty$ decreases with $k_1$, with a discontinuity when the number of lockdowns changes. After the mentioned $k_1$, $R_\infty$ increases, meaning that lower restrictiveness is worse in that regime. 

Meanwhile, for scale-free networks, $R_\infty$ monotonically decreases when successive lockdowns are allowed. As in the case of regular networks, when the number of lockdowns increases, $R_\infty$ has a discontinuity and is significantly higher. However, in the case of scale-free networks, $R_\infty$ is always higher when the lockdown is repeated, \textit{i.e.}, after the orange and red curves in Figure~\ref{R_infinity_b} have a discontinuity, they never get lower than the values they attain before said discontinuity. For regular networks, $R_\infty$ is in the same range regardless of the number of lockdowns. This difference is related to the mentioned ongoing pattern of scale-free networks having an extra wave of infections after the last lockdown ends, and thus will also be discussed in Section~\ref{section:discussion}.

\subsection{Cyclically repeating social distancing measures}\label{section:cycle}

In this section, we introduce a model that incorporates cyclic social distancing measures. As in the previous sections, when the total number of infected individuals, $I_\text{tot}$, exceeds a certain threshold, $\tau$, social distancing measures start. However, in this model, social distancing measures are implemented for a fixed duration of $d_1$ days, followed by a “rest period” of $d_2$ days. If $I_\text{tot}$ still exceeds $\tau$ at the end of the rest period, a new lockdown is initiated for $d_1$ days. This process is repeated cyclically, until $I_\text{tot}<\tau$ when a new lockdown would start at the end of a rest period.

\begin{figure}[ht]\centering
 \begin{subfigure}{.49\textwidth} \centering
 \includegraphics[width=\linewidth]{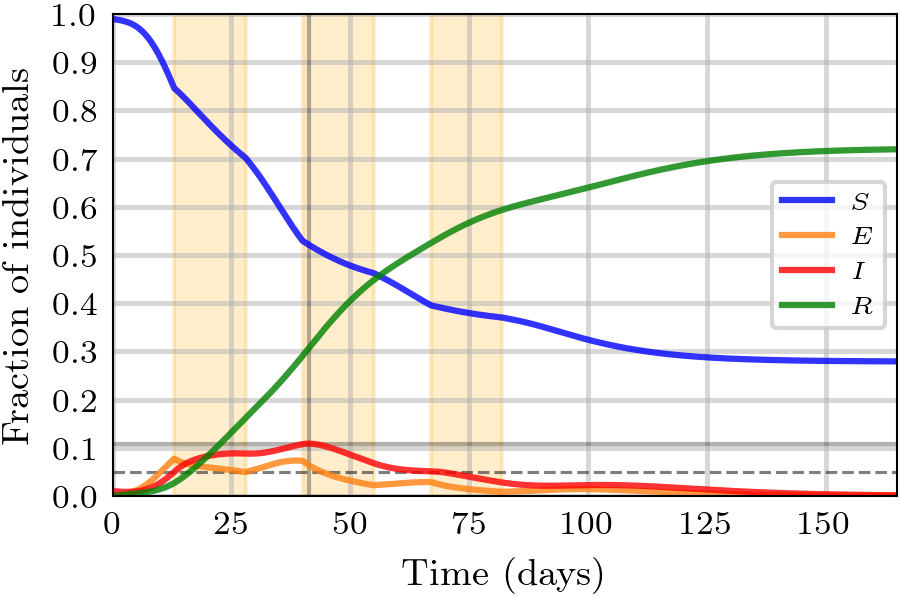}
\caption{$d_1=15, d_2=12$}
\end{subfigure}
\begin{subfigure}{.49\textwidth} \centering
\includegraphics[width=\linewidth]{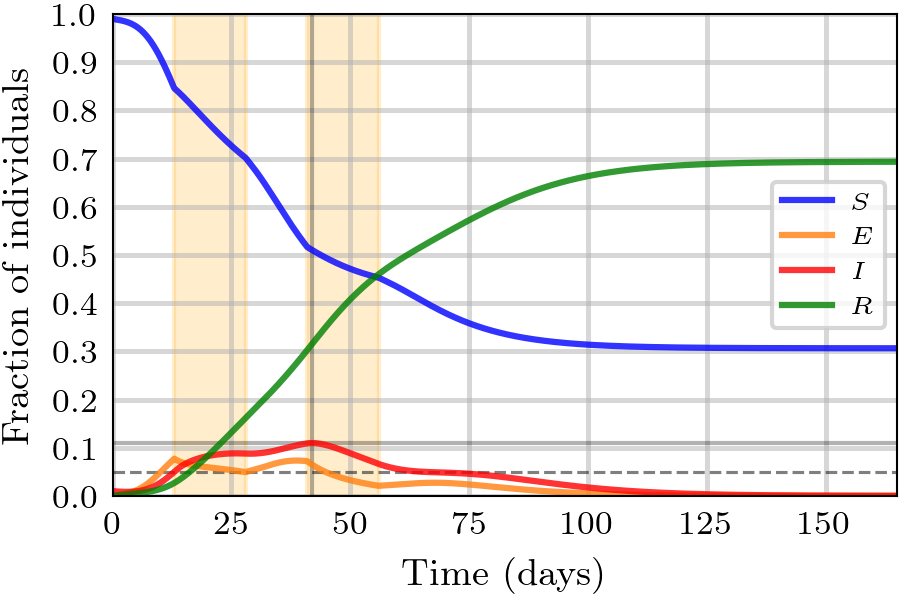}
\caption{$d_1=15, d_2=13$}
\end{subfigure}
\caption{Epidemic processes using cyclical lockdown strategies with a fixed duration and rest periods. The dashed line represents the threshold value $\tau=0.05$. }
 \label{cyclic_example}
\end{figure}

\begin{figure}[th]\centering
 \begin{subfigure}{.49\textwidth} \centering
 \includegraphics[width=\linewidth]{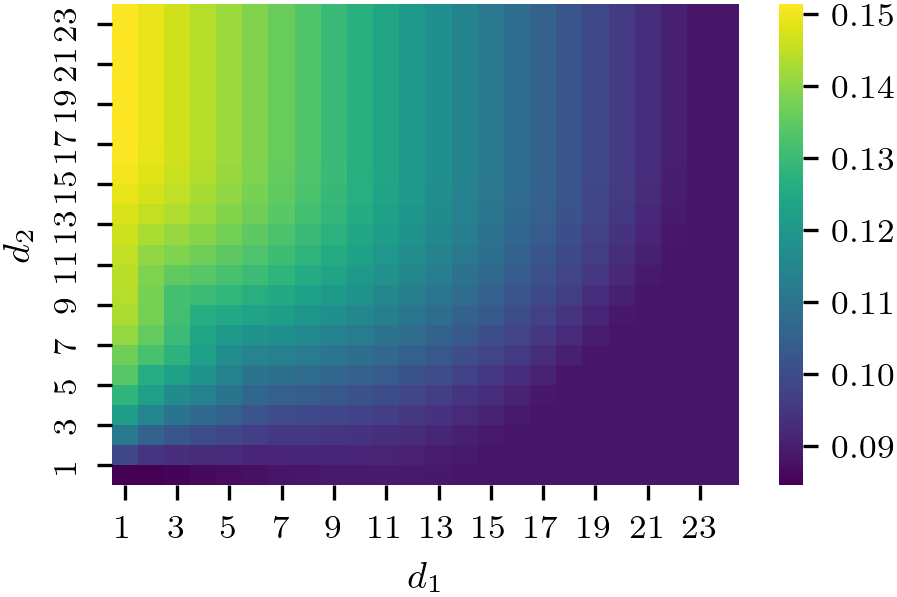}
\caption{$I_\text{max}$}
\end{subfigure}
\begin{subfigure}{.49\textwidth} \centering
\includegraphics[width=\linewidth]{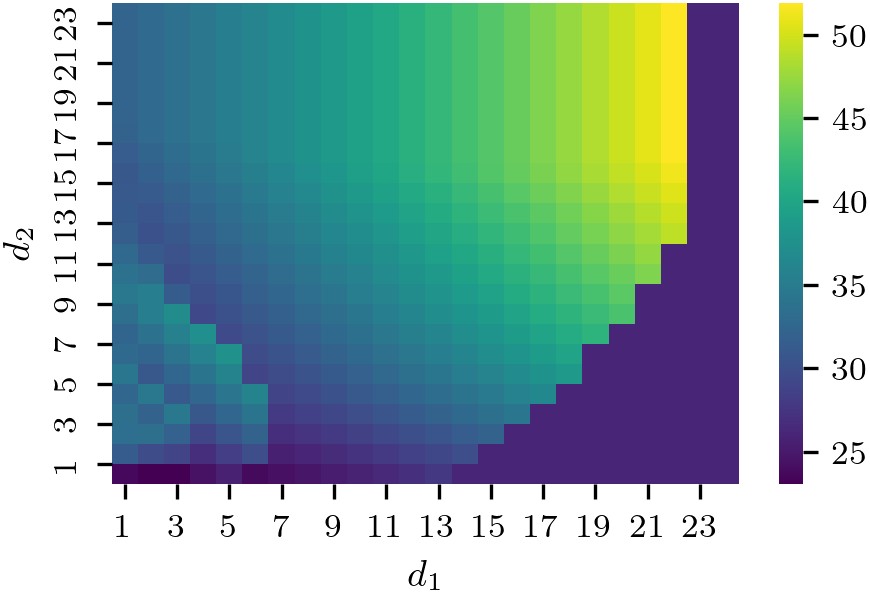}
\caption{$t(I_\text{max})$}
\end{subfigure}
\caption{Heatmaps of $I_\text{max}$ and $t(I_\text{max})$ as a function of both $d_1$ and $d_2$. }
 \label{cyclic_observables}
\end{figure}

We focus solely on scale-free networks and set $\gamma_0=2.5$, $\gamma_1=3$, and $\tau=0.05$. Examples of this process are illustrated in Fig.\ref{cyclic_example}. Heatmaps of $I_\text{max}$ and $t(I_\text{max})$ as a function of both $d_1$ and $d_2$ are shown in Figure~\ref{cyclic_observables}.

For $14\leq d_1\leq22$, there are certain values of $d_2$ for which both $I_\text{max}$ and $t(I_\text{max})$ reach their minimum values, while for $d_1\geq 23$ the same minimum value is reached $\forall d_2$. This occurs because for scale-free networks with $\gamma_0=2.5$, the threshold $I_\text{tot}=0.05$ is surpassed at $t=12.98$, and for an uninterrupted lockdown with $\gamma_1=3$, the maximum is $I_\text{max}=0.088$ at $t(I_\text{max})=26.01$, as shown in Figure~\ref{observables_norepe_d7}. Thus, when the lockdown is not interrupted, after $\Delta t=13.03$ days $I_\text{tot}(t)$ reaches this peak, which corresponds to the darkest region of the heatmaps in Figure~\ref{cyclic_observables}. We will refer to this peak as the “uninterrupted peak” from here on.

Therefore, if $d_1\geq14$, the curve reaches this peak. However if the lockdown ends immediately after the peak, two or more days without social distancing will result in a sudden increase in $I_\text{tot}$, surpassing the uninterrupted peak. For $d_1=15$, as two days pass between the uninterrupted peak and the end of the lockdown, having $d_2=2$ days without isolation is not enough for the maximum to increase, but $d_2\geq3$ will produce a higher peak, and so on. This phenomenon is illustrated in Figure~\ref{cyclic_example}, which has $d_1=15$ and $d_2>2$. The longer the time between the uninterrupted peak and the end of the lockdown, the higher the $d_2$ required to reach a higher $I_\text{max}$. When $d_1\geq23$, \textit{i.e.} when the lockdown is maintained for at least $10$ days after the uninterrupted peak, it is guaranteed that the second peak will not surpass the first one.

\begin{figure}[th]\centering
 \includegraphics[width=.6\linewidth]{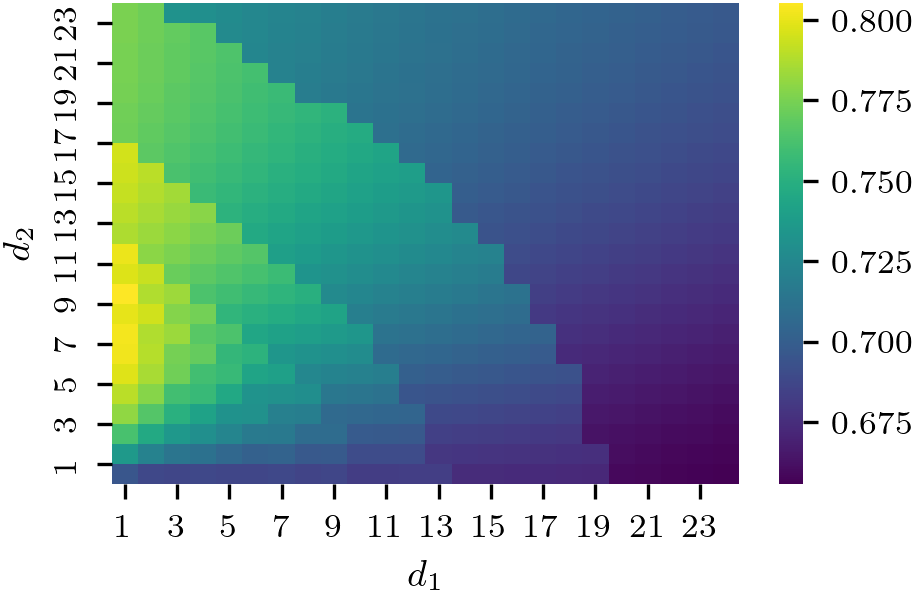}
 \caption{Heatmap of $R_\infty$ as a function of both $d_1$ and $d_2$.}
 \label{cyclic_Rinf}
\end{figure}

Figure~\ref{cyclic_observables} also shows that the highest $I_\text{max}$ is attained at the lowest $d_1$ and highest $d_2$, as expected when the lockdown periods are short and the periods without lockdown are long. However, the behavior of $R_\infty$, shown in Figure~\ref{cyclic_Rinf}, is not the same: for any given $d_1$, the lowest $R_\infty$ is attained for higher $d_2$. Therefore, having longer resting periods can actually decrease the final number of infections. 

This can be explained by noting that there are discontinuities in the heatmap of $R_\infty$, which correspond to the values of $d_1$ and $d_2$ for which increasing one of them will result in a change in the number of lockdown measures. For example, Figure~\ref{cyclic_example} shows that when $d_1=15$ and $d_2=12$ there are $3$ lockdowns while increasing $d_2$ to $13$ reduces the number of lockdowns to $2$, and the curve for $R_\text{tot}(t)$ tends to a lower value in the latter case. This behavior follows the same pattern encountered in previous sections, where we observed that having more lockdown measures in the case of scale-free networks increases $R_\infty$. Further discussion of this topic will be provided in Section~\ref{section:discussion}.

\newpage
\subsection{Social distancing strategies with complex degree distributions}\label{section:bimodal}

As the $K$-SEIR model works on arbitrary networks (as long as certain basic conditions are met~\cite{moreno2002}), it allows for more complex social behaviors to be represented in the degree distribution $\pi(k)$. In this section, we will take an initial approach to include the “essential workers” in the model: individuals with a high degree that must continue with their normal daily activities even during a lockdown. Thus, their degree will not change during the preventive measure, and this fact should be reflected in the new degree distribution. 

To construct this degree distribution, we will assume that in normal conditions $\pi^\mathrm i(k)$ will be that of a scale-free network, and during the lockdown, $\pi^\mathrm f(k)$ will be a combination of a power-law and a Gaussian function centered at a certain degree $k=k_\text{peak}$. To find the relative weights of each function, so that the bimodal distribution maintains the probability of the desired degree, we define
\begin{equation}
\pi_1(k)=\alpha_1 k^{-\gamma_1},\quad\pi_2(k)=\alpha_2k^{-\gamma_2},\quad
G(k)=\displaystyle \exp\left(-\frac{(k-k_\text {peak})^2}{2\sigma_\text{peak}^2}\right)
\end{equation}
\noindent then the degree distribution before and after the lockdown starts, $\pi^\mathrm i(k)$ and $\pi^\mathrm f(k)$ respectively, are
\begin{equation}
\pi^\text i(k)=\pi_1(k)\quad, \quad\pi^\text f(k)=A[\pi_2(k)+\beta\ G(k)]
\end{equation}
Remembering that $\pi^\mathrm f(k)$ must be normalized and $\pi^\textrm i(k_\text{peak})=\pi^\textrm f(k_\text{peak})$, 
\begin{equation}
\left\{\begin{matrix}
A=\left[\overbrace{\sum\pi_2(k)}^1+\beta\sum G(k)\right]^{-1}=\left[1+\beta\sum G(k)\right]^{-1}\\
\pi_1(k_\text{peak}) = A[\pi_2(k_\text{peak})+\beta \underbrace{G(k_\text{peak})}_1]= A[\pi_2(k_\text{peak})+\beta]
\end{matrix}\right.
\end{equation}
Finally, solving for the relative weight $\beta$,
\begin{equation}
\beta = \frac{1-\pi_2(k_\text{peak})/\pi_1(k_\text{peak})}{1/\pi_1(k_\text{peak})-\sum G(k)}=\frac{\pi_1(k_\text{peak})-\pi_2(k_\text{peak})}{1-\pi_1(k_\text{peak})\cdot\sum G(k)}
\end{equation}

\begin{figure}[thb]\centering
 \begin{subfigure}{.49\textwidth} \centering
 \includegraphics[width=\linewidth]{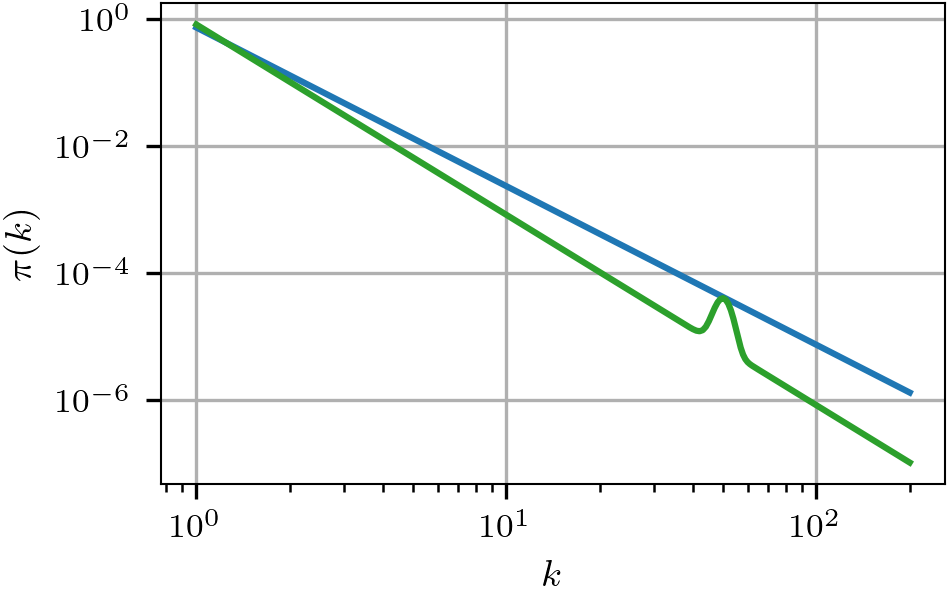}
\caption{$k_\text{peak}=50$.}
\label{bimodal_ejemplo_a}
\end{subfigure}
\begin{subfigure}{.49\textwidth} \centering
\includegraphics[width=\linewidth]{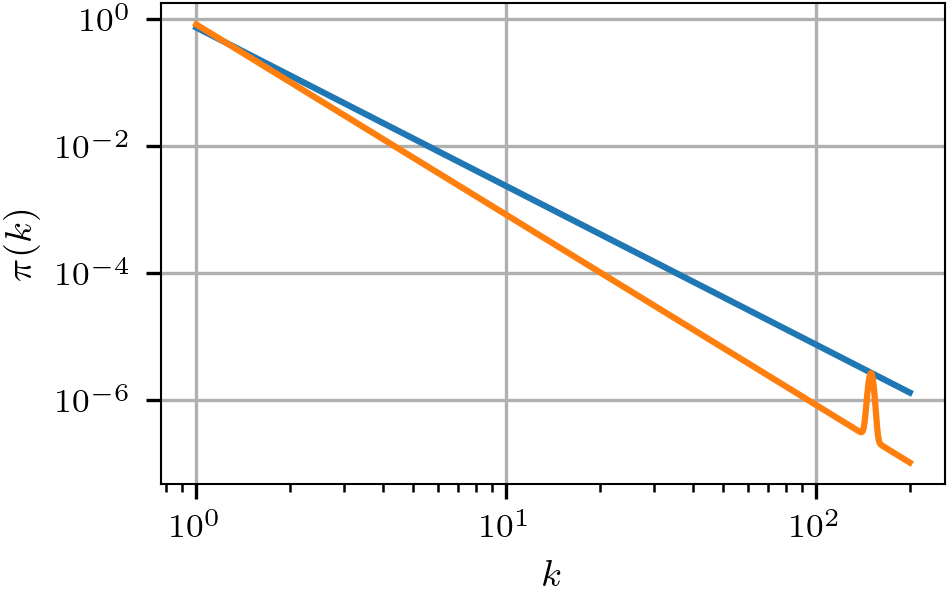}
\caption{$k_\text{peak}=150$.}
\label{bimodal_ejemplo_b}
\end{subfigure}
\caption{The curve in blue represents the initial distribution $\pi^\text i(k)=\alpha_1 k^{-2.5}$ while the orange and green ones are bimodal distributions $\pi^\text f(k)=A\left[\alpha_2 k^{-3}+\beta \exp\left(-\frac{(k-k_\text{peak})^2}{18}\right)\right]$.}
 \label{bimodal_example}
\end{figure}

We will fix $\gamma_0=2.5$, $\gamma_1=3$, and $\sigma_\mathrm{peak}=3$. Examples of the resulting $\pi^\mathrm f(k)$ are shown in Figure~\ref{bimodal_example}. 

\begin{figure}[th]\centering
 \includegraphics[width=.7\linewidth]{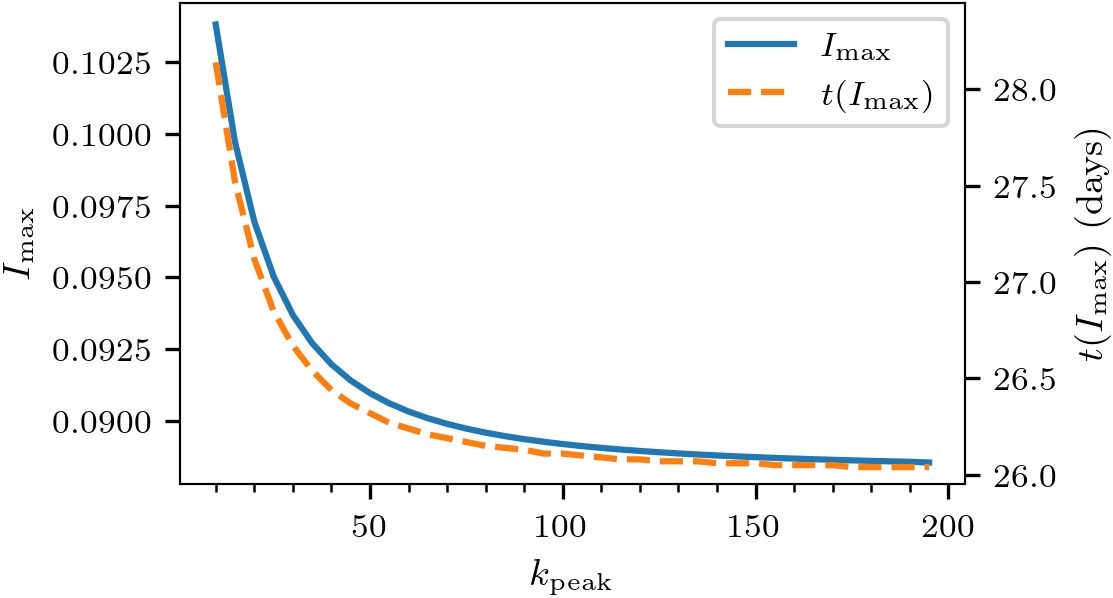}
 \caption{$I_\text{max}$ and $t(I_\text{max})$ as a function of $k_\text{peak}$ for bimodal distributions.}
 \label{bimodal_observables}
\end{figure}

In this section, we aim to address the question of whether it is more effective from an epidemiological standpoint to have a larger number of individuals with a lower degree or a smaller number of individuals with a higher degree. For instance, when $k_\text{peak}=50$, the fraction of individuals that do not isolate is proportional to $50^{-2}$, and when $k_\text{peak}=150$, it will be proportional to $150^{-2}$. To explore this question, we vary the value of $k_\text{peak}$, and compare the resulting values of $I_\text{max}$ and $t(I_\text{max})$ for each case. We present the results of our analysis in Figure~\ref{bimodal_observables}, where we consider lockdowns that do not repeat and have parameters $d=7$ and $\tau=0.05$ as in Section~\ref{section:non_repeating}.

As $I_\text{max}$ decreases with $k_\mathrm{peak}$, this suggests that better epidemiological results are obtained when fewer individuals have a higher degree. This is a good example of how certain complex behaviors can be included in the $K$-SEIR model by creatively adapting the degree distribution as the situation demands it.

\section{Discussion}\label{section:discussion}
Throughout Sections.~\ref{section:non_repeating}, \ref{section:succesive}, and \ref{section:cycle}, a recurring pattern arose while analyzing the differences between the results for regular and scale-free networks. Firstly, we noted that when there is only one lockdown, in regular networks there is a range of $k_1$ for which only one wave of infections occurs, while in scale-free networks there are always two waves. This is more notable in the case of $d=70$, as illustrated in Figure~\ref{infected_norepe_d70}. Furthermore, when successive lockdowns are implemented, we once again see that in scale-free networks, there is always an additional wave after the last lockdown ends. 

\begin{figure}[th]\centering
 \includegraphics[width=.6\linewidth]{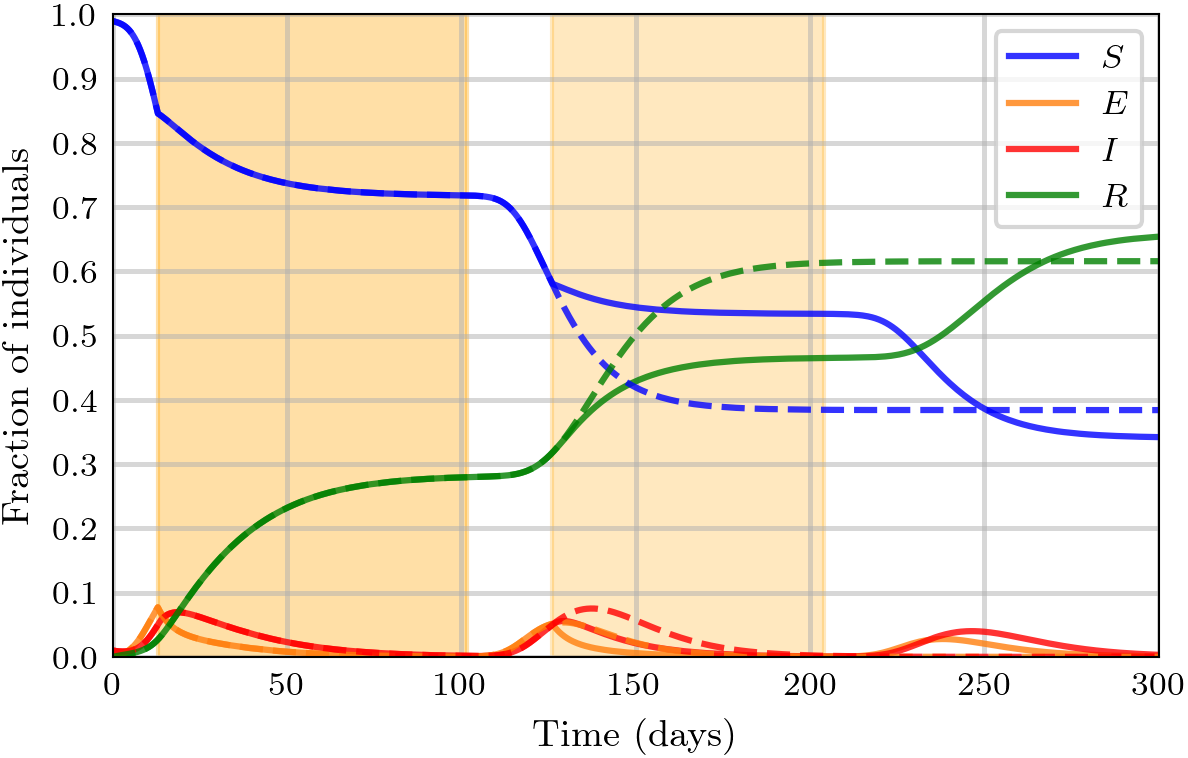}
 \caption{Evolution of the total fraction of individuals in the epidemiological compartments in a scale-free network. Solid lines correspond to the successive lockdown strategy, while dashed lines correspond to the non-repeating social distancing strategy. Highlighted in orange are the times for which the lockdown is in place.}
\label{overlap_gamma1_3.5_d70}
\end{figure}

While this difference does not affect $I_\text{max}$ or $t(I_\text{max})$, it impacts on the behavior of $R_\infty$, as seen in Figure~\ref{R_infinity}. In that Figure, we see that for regular networks the value of $R_\infty$ is in the same range regardless of the number of lockdowns, while for scale-free networks it is clear that a higher number of lockdown measures implies a higher value of $R_\infty$. To analyze this behavior for scale-free networks, Figure~\ref{overlap_gamma1_3.5_d70} shows the evolution of the total fraction of individuals in the epidemiological compartments for repeating and non-repeating lockdown strategies simultaneously. The parameters are set to $\gamma_0=2.5$, $\gamma_1=3.5$, $\tau=0.05$, and $d=70$, as this case allows us to observe each wave more clearly and tell them apart. 

\begin{figure}[thb]\centering
\begin{subfigure}{.32\textwidth} \centering
\includegraphics[width=\linewidth]{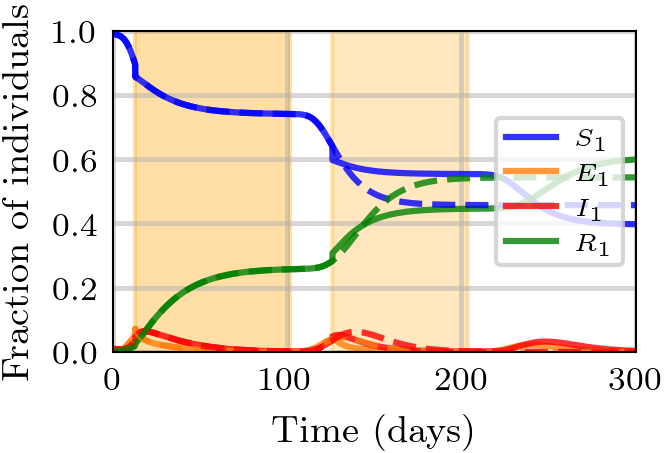}
\caption{$k=1$}
\end{subfigure}
\begin{subfigure}{.32\textwidth} \centering
\includegraphics[width=\linewidth]{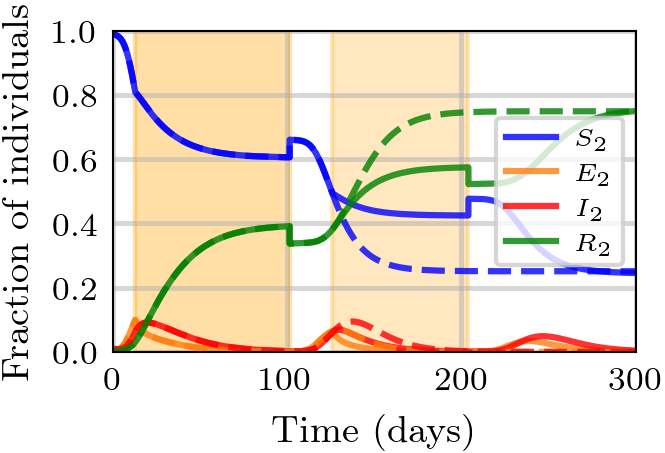}
\caption{$k=2$}
\end{subfigure}
\begin{subfigure}{.32\textwidth} \centering
\includegraphics[width=\linewidth]{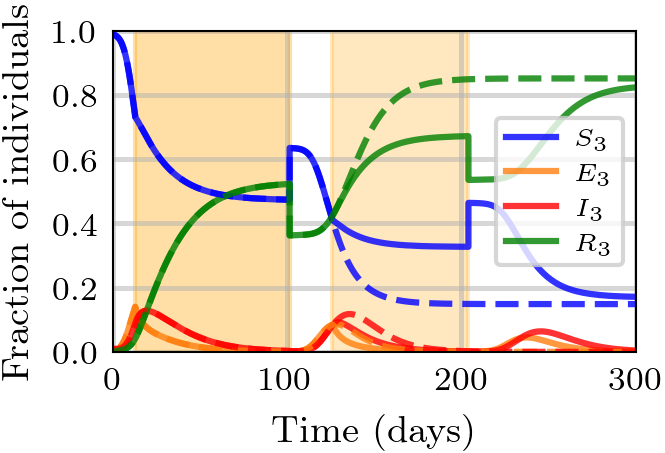}
\caption{$k=3$}
\end{subfigure}
\begin{subfigure}{.32\textwidth} \centering
\includegraphics[width=\linewidth]{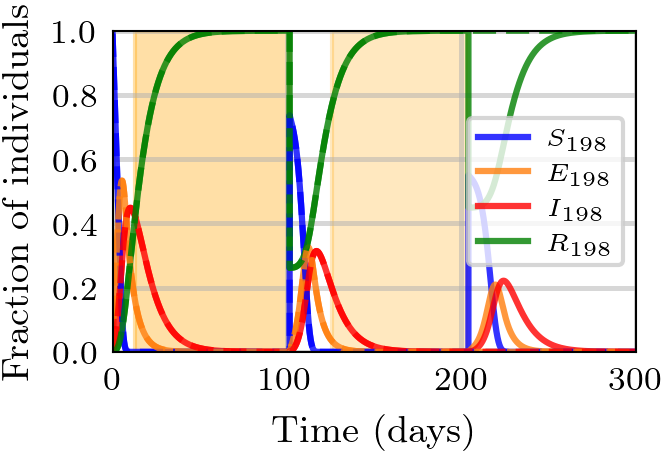}
\caption{$k=198$}
\end{subfigure}
\begin{subfigure}{.32\textwidth} \centering
\includegraphics[width=\linewidth]{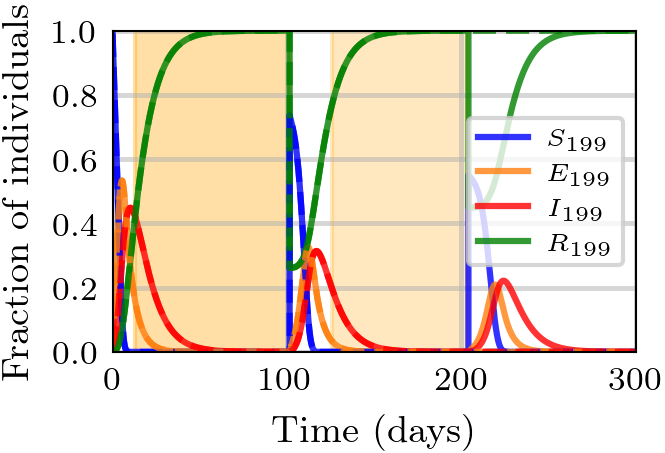}
\caption{$k=199$}
\end{subfigure}
\begin{subfigure}{.32\textwidth} \centering
\includegraphics[width=\linewidth]{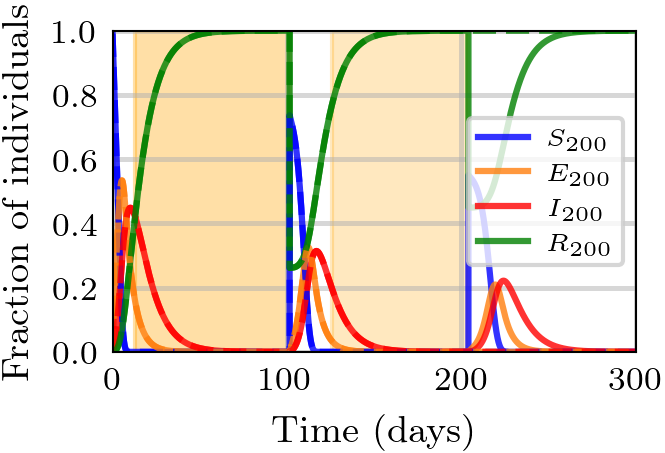}
\caption{$k=200$}
\end{subfigure}
\caption{Evolution of the fraction of individuals in the sub-compartments corresponding to the lowest and highest degrees of the processes shown in Figure~\ref{overlap_gamma1_3.5_d70}.}
 \label{sub-compartments}
\end{figure}

As previously mentioned, the repeated lockdown scenario leads to a third wave of infections due to susceptible individuals returning to their original number of daily contacts. It is during this third wave that the curve of $R_\text{tot}(t)$ surpasses the one for the case of the single lockdown. To better see the dynamics behind this, Figure~\ref{sub-compartments} shows the evolution of the sub-compartments corresponding to the three lowest degrees and three highest ones. Notice that until now, we have always shown the evolution of each compartment calculated by summing the fraction of individuals in its sub-compartments according to Eq.~(\ref{rho_tot}).

We can see that when a lockdown measure ends, the sub-compartments corresponding to the higher degrees “receive” a higher fraction of susceptible individuals, that are then quickly infected because of the high exposition they get. The individuals that extremely reduced their degree act as the “fuel” for the additional wave when they are mixed after the preventive measure. In other words, the sub-compartments with lower degrees serve as a “reservoir” of susceptible individuals that are then scattered among the higher degrees when the lockdown ends. This can not occur in regular networks, as all of the individuals have the same degree: even while isolated, there is no group of individuals that get infected more than any other group, and in consequence, this “reservoir” effect can not take place on regular networks. 

\section{Conclusions}\label{section:conclusions}
We have developed a mean-field mesoscopic epidemiology model that offers the analytical advantages of macroscopic compartmental models while considering the social structure of the population. This model is well-suited for simulating the impact of social distancing measures by incorporating changes to the social contact patterns, included in the model via the degree distribution of the population.

Our study found that the social structure of the population, as represented by the degree distribution, plays a crucial role in the effectiveness of social distancing measures. In the model, the subcompartments of individuals with lower degrees act as a "reservoir" of susceptible individuals that are then scattered among the higher degrees when the lockdown ends. This effect causes a wave of infections after each lockdown measure. It is important to note that this effect does not occur in regular networks because all individuals have the same degree, and for the same reason this effect can not be found in classical compartmental models that consider the population to be completely homogeneous. 

Notably, our model can predict successive waves of infections without the need to account for reinfections. This feature allows it to qualitatively reproduce the wave patterns observed across many countries during the COVID-19 pandemic, including cases in which the second wave has a higher peak of infections compared to the first one~\cite{owidcoronavirus}.

The model is highly flexible and can implement various social distancing strategies, adjust the threshold for triggering a lockdown measure, its duration, or whether it occurs once or repeatedly. The model can also incorporate complex social behaviors by adapting the degree distribution accordingly, such as including essential workers.

Future developments of the model could include segmenting the population into different groups based on age or vulnerability to the infection, incorporating the possibility of reinfection and exploring its impact on successive waves, or evaluating the effectiveness of vaccination programs by prioritizing individuals with different degrees.


\bibliographystyle{elsarticle-num}

\end{document}


\newcommand*{\hwplotB}{\raisebox{3pt}{\tikz{\draw[red,dashed,line 
width=3.2pt](0,0) -- 
(5mm,0);}}}

\newrobustcmd*{\mydiamond}[1]{\tikz{\filldraw[black,fill=#1] (0.05,0) -- 
(0.2cm,0.2cm) -- (0.35cm,0) -- (0.2cm,-0.2cm) -- (0.05,0);}}

\newrobustcmd*{\mytriangleright}[1]{\tikz{\filldraw[black,fill=#1] (0,0.2cm) -- 
(0.3cm,0) -- (0,-0.2cm) -- (0,0.2cm);}}

\newrobustcmd*{\mytriangleup}[1]{\tikz{\filldraw[black,fill=#1] (0,0.3cm) 
-- (0.2cm,0) -- (-0.2cm,0) -- (0,0.3cm);}}

\newrobustcmd*{\mytriangleleft}[1]{\tikz{\filldraw[black,fill=#1] (0,0.2cm) -- 
(-0.3cm,0) -- (0,-0.2cm) -- (0,0.2cm);}}
\definecolor{Blue}{cmyk}{1.,1.,0,0} 

\begin{frontmatter}

\title{Implementing lockdown measures in epidemic outrbreaks through mean-field models considering the social structure} 

\address[IB]{Instituto Balseiro, Universidad Nacional de Cuyo}
\address[CONICET]{Consejo Nacional de Investigaciones Científicas y Técnicas (CONICET), Bariloche, Argentina}
\address[CNEA]{Gerencia de Física, Centro Atómico Bariloche, Comisión Nacional de Energía Atómica }

\author[IB,CONICET]{E.A.~Rozan}
\author[CONICET,CNEA]{S.~Bouzat}
\author[IB,CONICET,CNEA]{M.N.~Kuperman}

\end{frontmatter}
{\begin{center}
\LARGE Supplementary Material
\end{center}}

In this supplementary material, we explore different values for the parameters of the $K$-SEIR model, when it follows its natural evolution without lockdown measures. The parameters used in the numerical simulations discussed in the main document are determined by the findings presented in this supplementary material. These findings will be presented separately for regular networks and scale-free networks.

\section{Regular networks}
For regular networks, the equations of the $K$-SEIR model are simplified. Considering that $\pi(k)=\delta_{k,k_0}$, the term of daily contagions is
\begin{equation}
r\,\frac{kS_k}{\langle k\rangle}\sum_{k'}I_{k'}(k'-1)\pi(k')= r\,\frac{kS_k}{k_0}\sum_{k'}I_{k'}(k'-1)\delta_{k',k_0}=r\,\frac{kS_k}{k_0} \ \ I_{k_0}(k_0-1)
\end{equation}
Of the original $4K$ equations, only the ones corresponding to the degree $k=k_0$ survive. In those equations,
\begin{equation}
r\,\frac{kS_k}{\langle k\rangle}\sum_{k'}I_{k'}(k'-1)\pi(k')= r\,S_{k_0} I_{k_0}(k_0-1)
\end{equation}

\noindent Thus, the 4 equations of the sub-compartments with $k=k_0$ are simplified to

\begin{equation}
\begin{NiceMatrix}[l]
\displaystyle\frac{\text d S_{k_0}}{\text d t} =&
\displaystyle r\,S_{k_0} I_{k_0}(k_0-1)\\[0.5cm]
\displaystyle \frac{\text d E_{k_0}}{\text d t} =&
\displaystyle \phantom{-}r\,S_{k_0} I_{k_0}(k_0-1) - \frac{E_{k_0}}{T_\text{inc}}\\[0.5cm]
\displaystyle \frac{\text d I_{k_0}}{\text d t} =&
\displaystyle  \frac{E_{k_0}}{T_\text{inc}}-\frac{I_{k_0}}{T_\text{inf}}\\[0.5cm]
\displaystyle\frac{\text d R_{k_0}}{\text d t} =&
\displaystyle  \frac{I_{k_0}}{T_\text{inf}}\end{NiceMatrix}
\label{regular_ecuaciones}
\end{equation}

The initial conditions are set to $E_{k_0}(0)=R_{k_0}(0)=0$, $I_{k_0}(0)=0.01$ and ${S_{k_0}(0)=0.99}$, as in the main article document. The incubation and infection times are $T_\text{inc}=5.2$ and $T_\text{inf}=14-T_\text{inc}=8,8$, respectively, as initially reported for COVID-19~\cite{early_transmission_dynamics}. 

\begin{figure}[ht]\centering
 \begin{subfigure}{.4\textwidth} \centering
 \includegraphics[width=\linewidth]{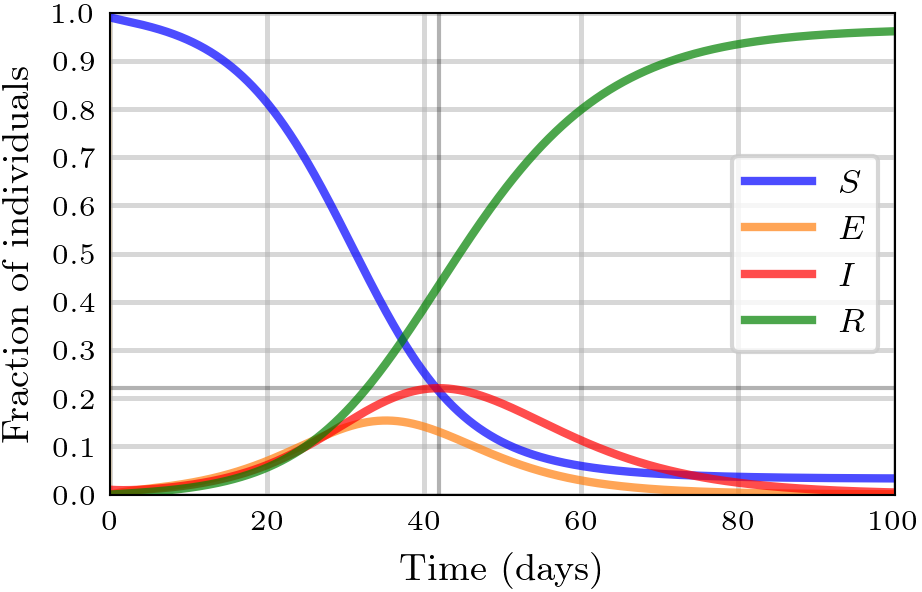}
 \caption{$k_0=5, r=0.1$}
 \end{subfigure}
  \begin{subfigure}{.4\textwidth} \centering
 \includegraphics[width=\linewidth]{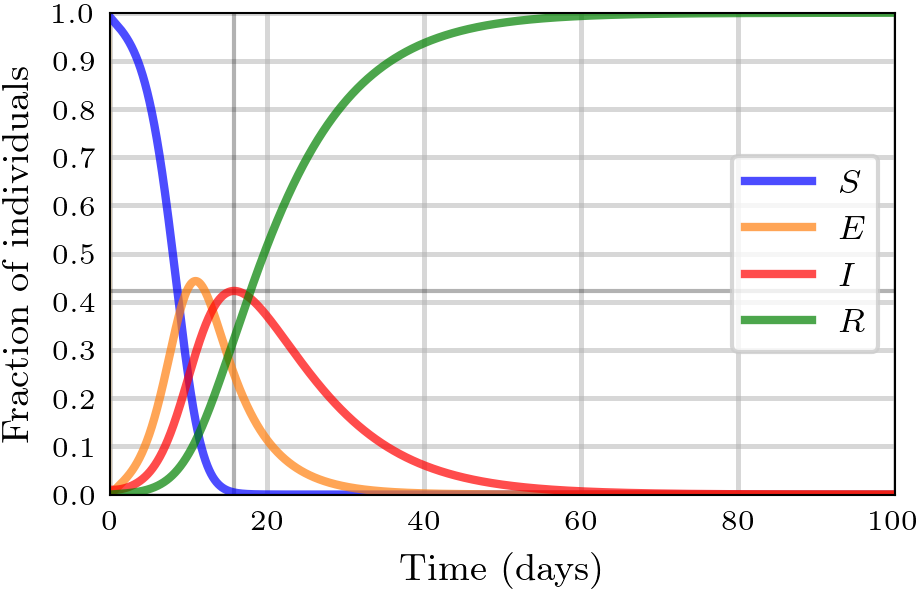}
 \caption{$k_0=20, r=0.1$ }
 \end{subfigure}
   \begin{subfigure}{.4\textwidth} \centering
 \includegraphics[width=\linewidth]{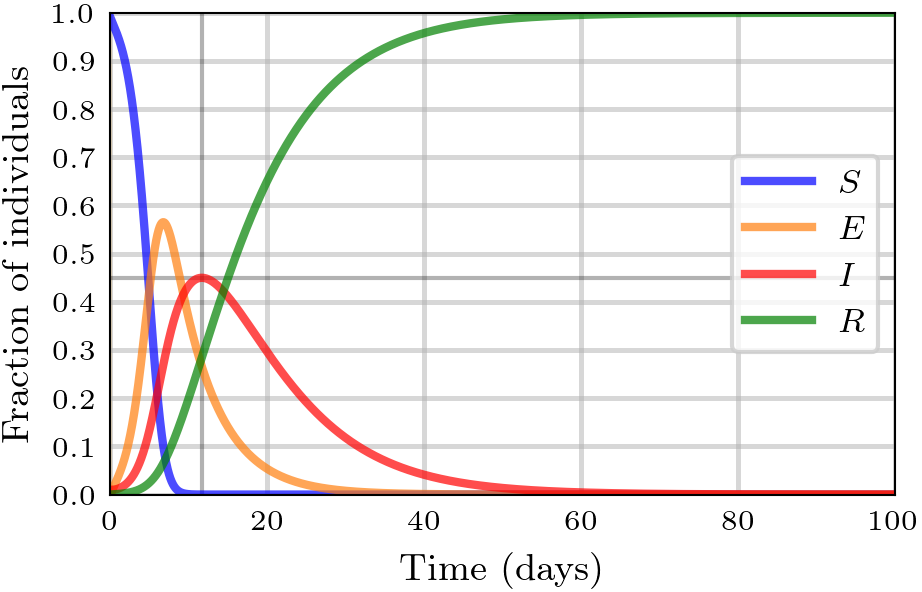}
 \caption{$k_0=5, r=1$ }
 \end{subfigure}
   \begin{subfigure}{.4\textwidth} \centering
 \includegraphics[width=\linewidth]{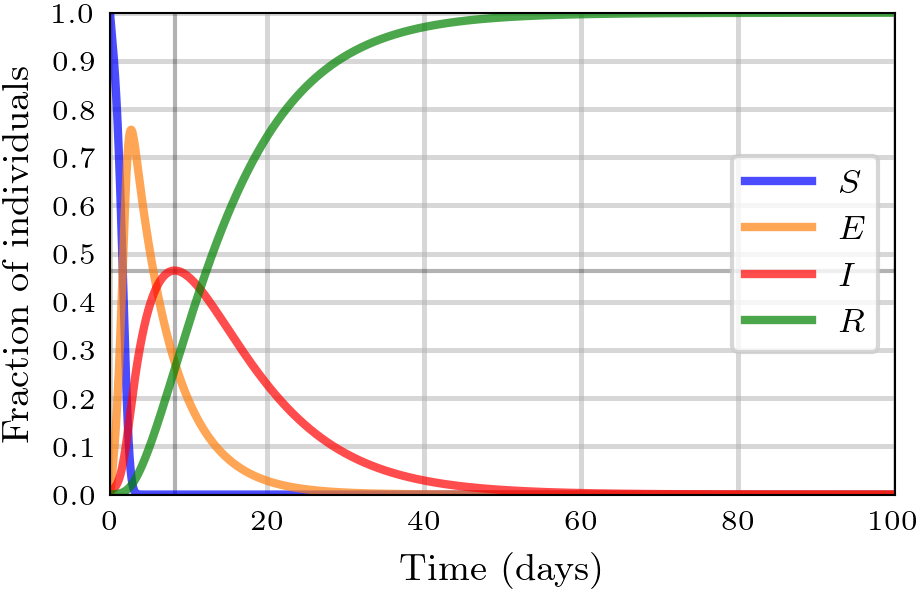}
 \caption{$k_0=20, r=1$}
 \end{subfigure}
 \caption{Time evolution of epidemic processes on regular networks.}
 \label{regular_fig1}
\end{figure}

To provide an initial overview of the system's evolution for different parameter sets, we show in Figure~\ref{regular_fig1} the evolution of the system for two values of $r$ and $k_0$. 

As expected, higher values of $r$ and $k_0$ lead to higher peaks of infections $I_\text{max}$ and a shorter time to reach the peak $t(I_\text{max})$. This is because an increase in the daily contacts of the population results in a quicker and easier propagation of the infection.

\begin{figure}[ht]\centering
 \begin{subfigure}{.49\textwidth} \centering
 \includegraphics[width=\linewidth]{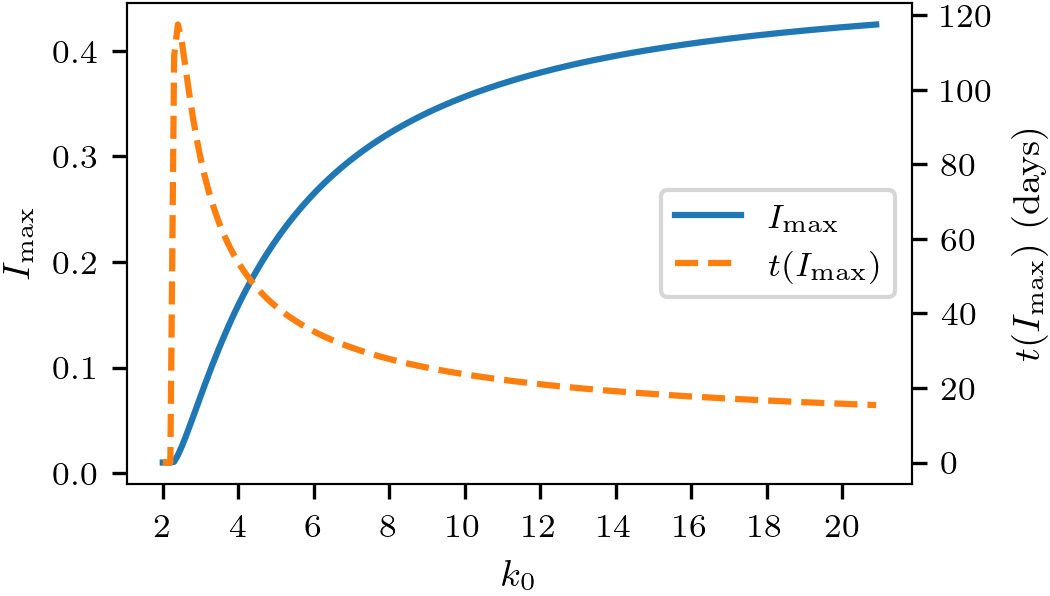}
 \caption{$r=0.1$}
  \label{observables_vs_k0_a}
 \end{subfigure}
  \begin{subfigure}{.49\textwidth} \centering
 \includegraphics[width=\linewidth]{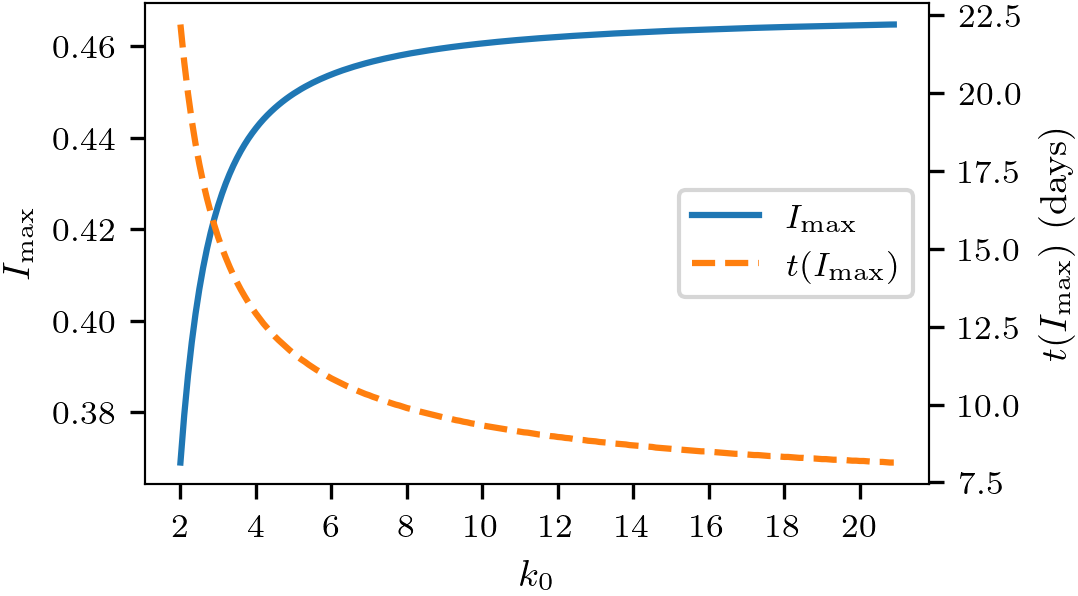}
 \caption{$r=1$}
   \label{observables_vs_k0_b}
 \end{subfigure}
 \caption{$I_\text{max}$ and $t(I_\text{max})$ as a function of $k_0$.}
 \label{observables_vs_k0}
\end{figure}

To see in detail how $I_\text{max}$ and $t(I_\text{max})$ depend on the connectivity of the individuals, Figure~\ref{observables_vs_k0} shows these quantities as a function of $k_0$ for the same values of $r$ shown in Figure~\ref{regular_fig1}. 

The height of the peak increases with $k_0$, while its time decreases. The most notable highlight of Figure~\ref{observables_vs_k0_a} is that from $k_0=2$ and $k_0=3$, there is a discontinuity in the curve of $t(I_\text{max})$ (colored in orange). In fact, for $k_0=2$ we see that $t(I_\text{max})$ and $I_\text{max}\simeq0$, which implies that the infection did not spread in the population and the maximum of $I(t)$ corresponds with the initial condition. In contrast, there is no $k_0$ with an analogous behavior in Figure~\ref{observables_vs_k0_b} in which $r$ is $10$ times higher.

\begin{figure}[th]\centering
 \begin{subfigure}{.49\textwidth} \centering
 \includegraphics[width=\linewidth]{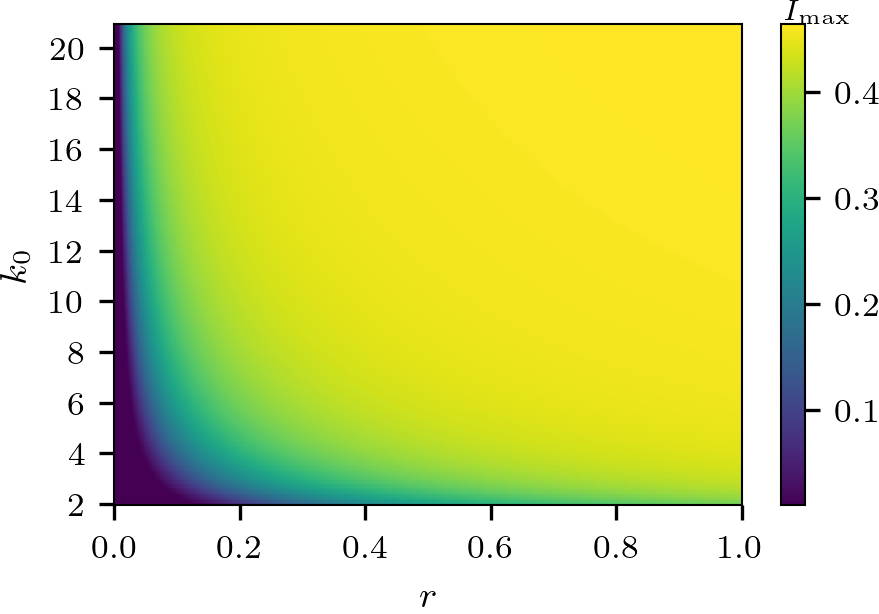}
 \caption{$I_\mathrm{max}$}
 \label{observables_vs_k0_r_a} 
 \end{subfigure}
  \begin{subfigure}{.49\textwidth} \centering
 \includegraphics[width=\linewidth]{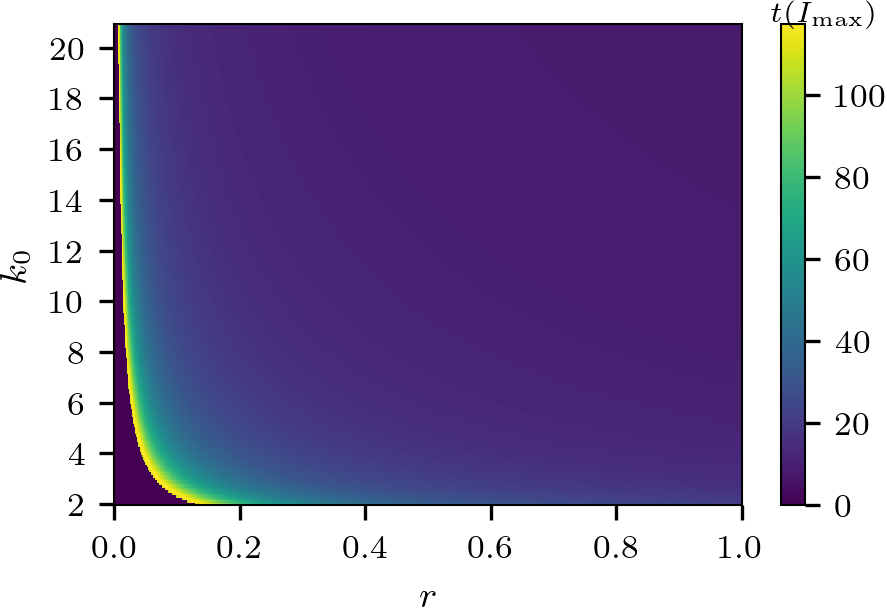}
 \caption{ $t(I_\mathrm{max})$}
 \label{observables_vs_k0_r_b}
 \end{subfigure}
 \caption{Heatmaps of $I_\text{max}$ and $t(I_\text{max})$ as a function of both $k_0$ and $r$.}
 \label{observables_vs_k0_r}
\end{figure}

This raises the question of when there is a critical value $k_0^\text{crit}$ for which the epidemic cannot spread, for certain values of $r$. To answer this question, we present heatmaps of $I_\text{max}$ and $t(I_\text{max})$ as a function of both $r$ and $k_0$ in Figure~\ref{observables_vs_k0_r}.

The existence of a critical value of $k_0$ becomes clear noting that the bottom-left region of Fig.\ref{observables_vs_k0_r_b} has $t(I_\text{max})=0$. This region is delimited by a curve with high times, which is colored in yellow, and represents the same discontinuity as that shown in Figure~\ref{observables_vs_k0_a} when the $t(I_\text{max})$ goes from $0$ to $\sim100$ at $k_0=k_0^\text{crit}$.

This behavior seems to be related to the basic reproduction number $R_0$ of the model. Based on the results of Ref.~\cite{SEIR_R0}, the expression of $R_0$ in this model can be written as follows:
\begin{equation}
R_0=rT_\text{inf}(k_0-1)
\end{equation}
\noindent and in the critical case of $R_0=1$, solving for $k_0^\text{crit}$ we get
\begin{equation}
k_0^\text{crit}(r)=1+\frac1{rT_\text{inf}}
\label{k0(r)}
\end{equation}

\begin{figure}[th]\centering
 \includegraphics[width=0.7\linewidth]{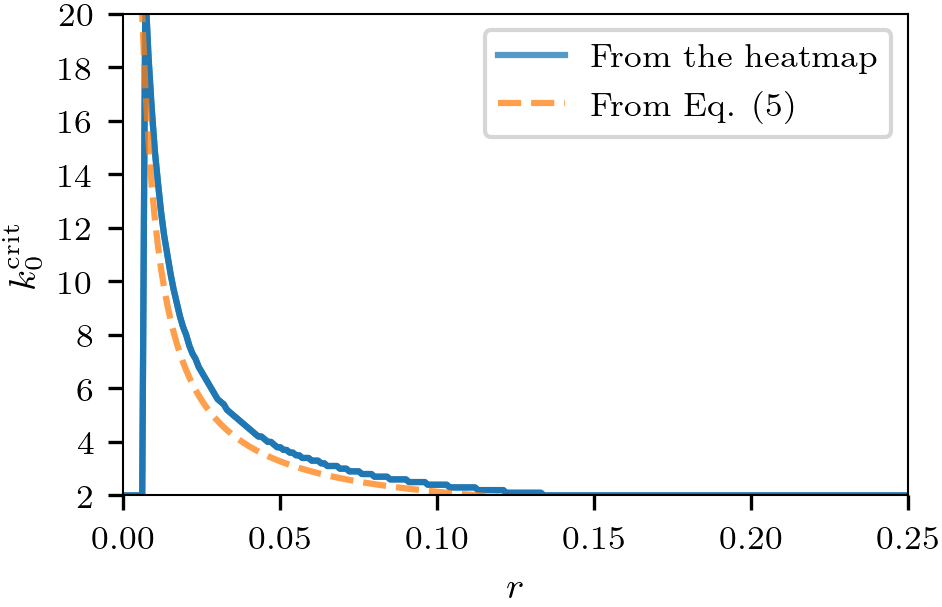}
 \caption{The blue curve represents the values of $k_0(r)$ for which $t(I_\text{max})$ attains a maximum in the heatmap of Figure~\ref{observables_vs_k0_r_b}. The orange curve is the plot of Eq.~\ref{k0(r)}.}
 \label{k0_crit}
\end{figure}

In Fig.\ref{k0_crit}, we present the values of $k_0^\text{crit}(r)$ obtained from the heatmap of Fig.\ref{observables_vs_k0_r_b} and the curve $k_0^\text{crit}$ from Eq.~(\ref{k0(r)}). Both plots coincide qualitatively and quantitatively in the explored range of $r$ and $k_0$. 

In the main article document, we chose $k_0=15$ as the starting amount of daily contacts of individuals, as in Ref.\cite{svoboda2022}. We also set the infection rate to $r=0.05$, which results in $R_0=6.16$. With these parameter values, $I_\text{max}$ is roughly $0.3$ as seen in Figure~\ref{observables_vs_k0_r_a}. We also set the threshold value to initiate an isolation measure to $\tau=0.1$ as in Ref.~\cite{svoboda2022}, which is approximately $I_\text{max}/3$.

\section{Scale-free networks}
\begin{figure}[ht]\centering
 \begin{subfigure}{.45\textwidth} \centering
 \includegraphics[width=\linewidth]{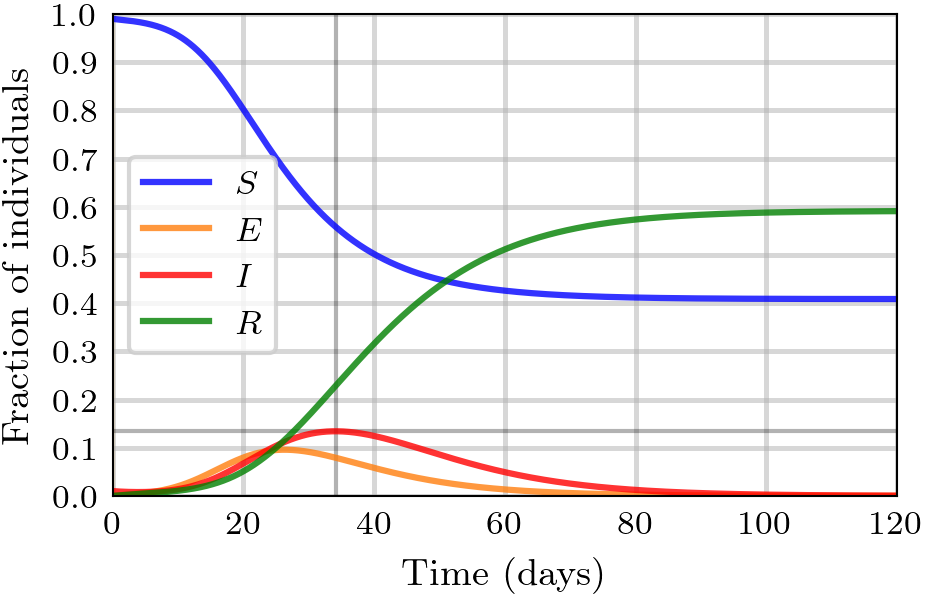}
 \caption{${K}=50,\gamma=2.5$}
 \end{subfigure}
  \begin{subfigure}{.45\textwidth} \centering
 \includegraphics[width=\linewidth]{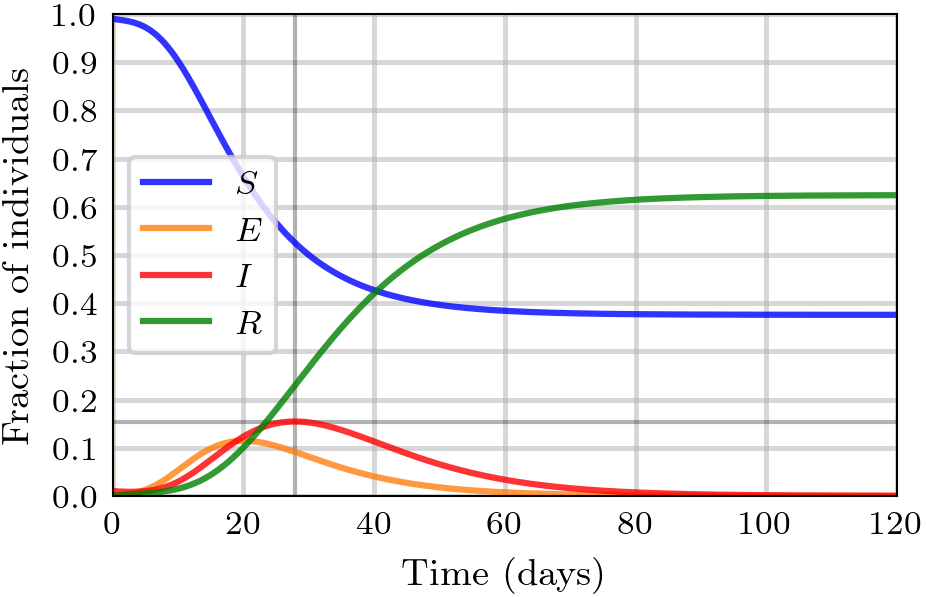}
 \caption{${K}=250,\gamma=2.5$}
 \end{subfigure}
  \begin{subfigure}{.45\textwidth} \centering
 \includegraphics[width=\linewidth]{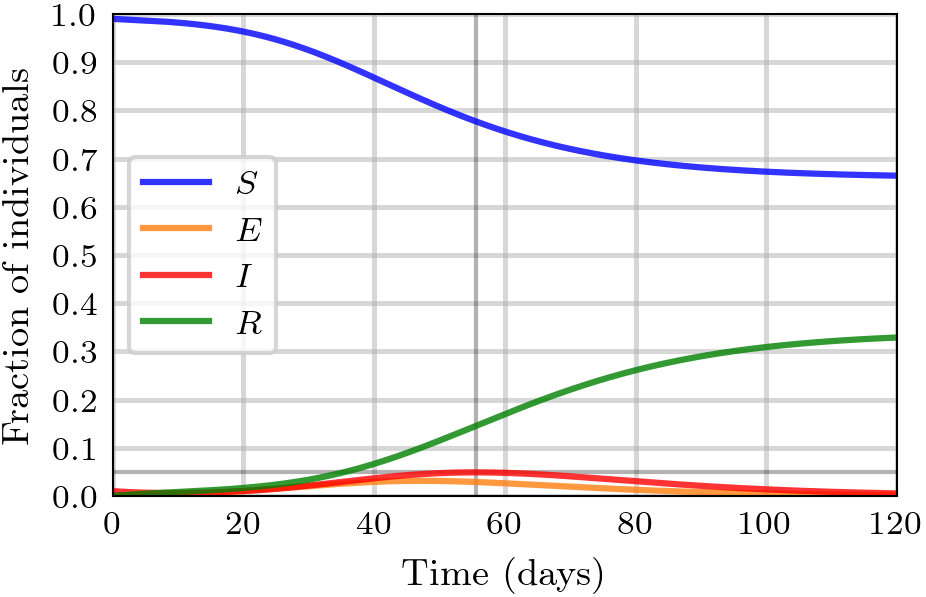}
 \caption{${K}=50,\gamma=3$}
 \end{subfigure}
  \begin{subfigure}{.45\textwidth} \centering
 \includegraphics[width=\linewidth]{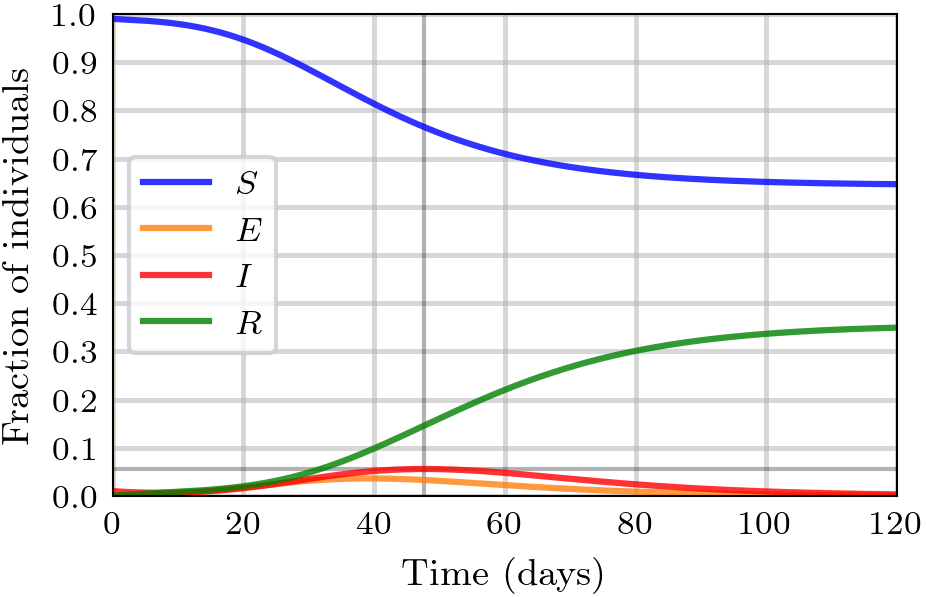}
 \caption{${K}=250,\gamma=3$}
 \end{subfigure}
 \caption{Evolution of the total fraction of individuals in the epidemiological compartments in scale-free networks with different parameters.}
 \label{KSEIR_examples}
\end{figure}

The equations of the $K$-SEIR model cannot be simplified in the case of scale-free networks. In this case, the degree distribution given by $\pi(k)=\alpha k^{-\gamma}\$ $ for $k=\{1,...,K\}$ has two parameters: $K$ and $\gamma$. Figure~\ref{KSEIR_examples} shows examples with two values of each parameter: in the top row $\gamma=2.5$ while in the bottom one $\gamma=3$, in the left column $K=50$ and in the right one $K=250$. The initial conditions are the same as in the main text: $E_k(0)=R_k(0)=0, {I_k(0)=0.01}$ and $S_k(0)=0.99\ \forall k$. The rest of the parameters are $r=0.2$, $T_\text{inc}=5.2$ and $T_\text{inf}=8.8$ days. 

\begin{figure}[ht]\centering
 \begin{subfigure}{.48\textwidth} \centering
 \includegraphics[width=\linewidth]{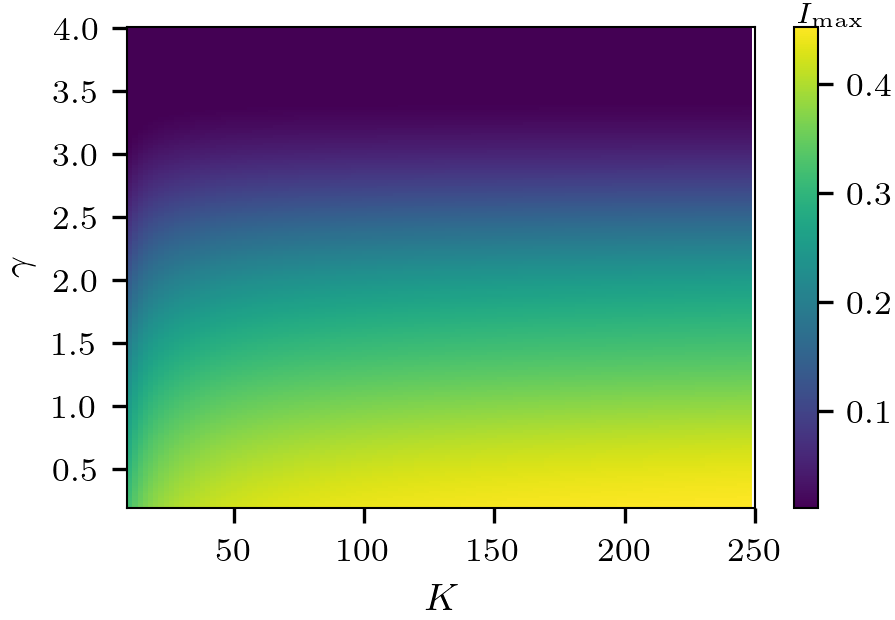}
 \caption{$I_\text{max}$}
 \end{subfigure}
  \begin{subfigure}{.48\textwidth} \centering
 \includegraphics[width=\linewidth]{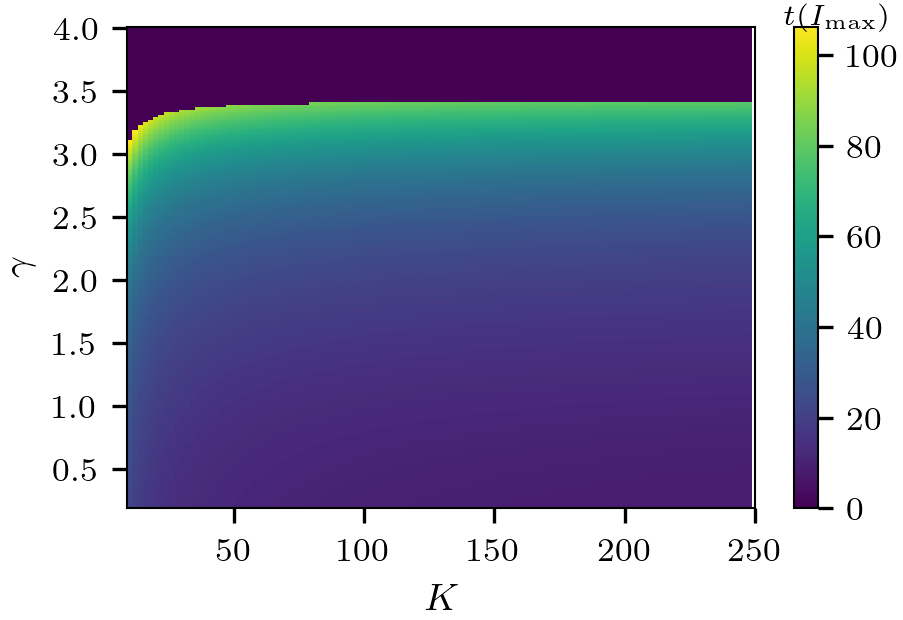}
 \caption{$ t(I_\text{max})$}
 \end{subfigure}
 \caption{Heatmaps of $I_\text{max}$ and $ t(I_\text{max})$ as a function of both ${K}$ and $\gamma$, the parameters of the degree distribution.}
 \label{KSEIR_heatmaps}
\end{figure}

\FloatBarrier
We can see that varying $\gamma$ has more noticeable effects on the results than changing $K$. Nonetheless, increasing $K$ and decreasing $\gamma$ result in a higher peak of infections $I_\text{max}$ and an earlier time of attainment $t(I_\text{max})$. Figure~\ref{KSEIR_heatmaps} presents heatmaps illustrating these quantities as a function of both $\gamma$ and $K$.

\FloatBarrier

\begin{figure}[ht]\centering
 \begin{subfigure}{.49\textwidth} \centering
 \includegraphics[width=\linewidth]{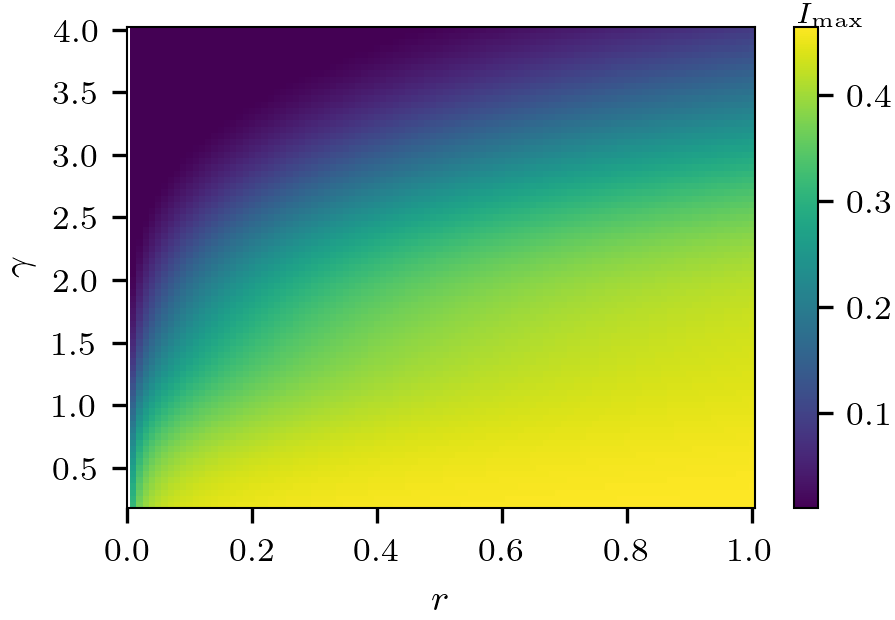}
 \caption{$I_\text{max}$}
  \label{KSEIR_heatmaps_r_a}
 \end{subfigure}
  \begin{subfigure}{.49\textwidth} \centering
 \includegraphics[width=\linewidth]{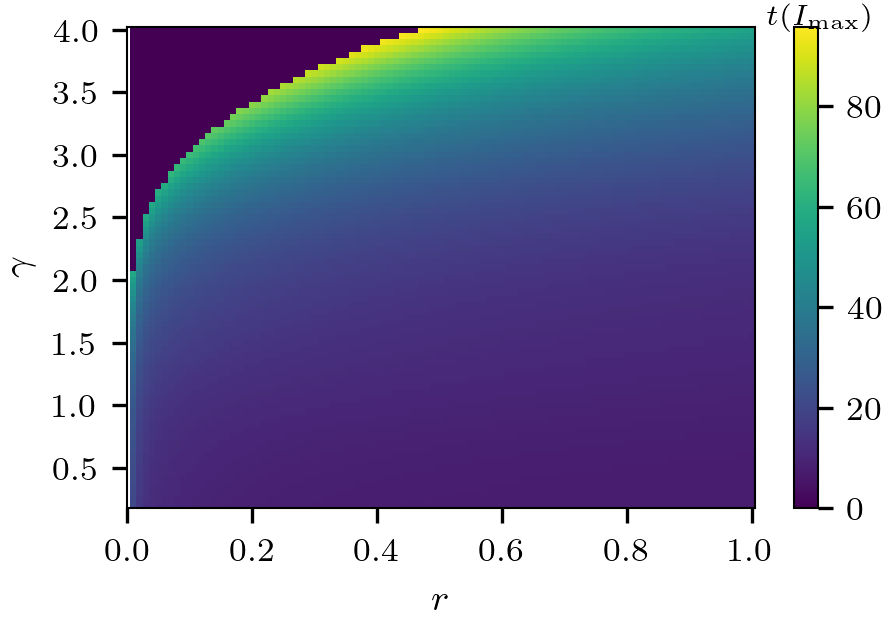}
 \caption{$ t(I_\text{max})$}
  \label{KSEIR_heatmaps_r_b}
 \end{subfigure}
 \caption{$I_\text{max}$ and $t(I_\text{max})$ as a function of $\gamma$ and $r$. ${K}$ is set to $200$. }
 \label{KSEIR_heatmaps_r}
\end{figure}

\begin{figure}[ht]\centering
 \includegraphics[width=0.65\linewidth]{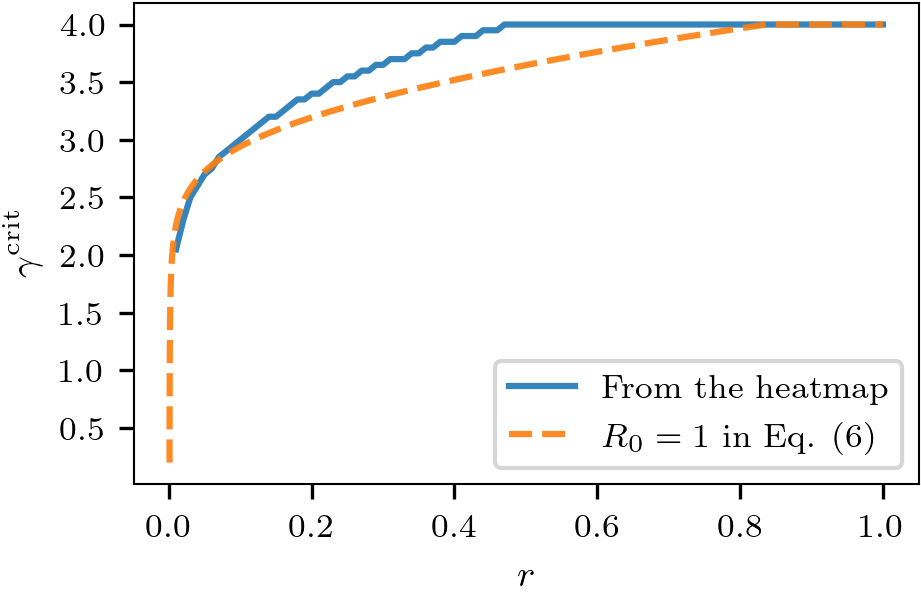}
 \caption{Values of $\gamma$ and $r$ for which $R_0=1$ according to Figure~\ref{KSEIR_heatmaps_r_b} and Eq.~(\ref{R0_KSEIR}).}
 \label{R0_curves}
\end{figure}

As with the preceding section, there exists a particular region in the parameter space where $t(I_\text{max})=0$, delineated by a curve of local maximum (colored in yellow). This area indicates the range in which the infection is incapable of disseminating throughout the population. Notably, for values of $K$ greater than or equal to 100, these outcomes are independent of $K$ but heavily reliant on the value of $\gamma$. Consequently, we set the maximum number of daily contacts an individual can have as $K=200$. To further scrutinize the threshold values of $\gamma$. Figure~\ref{KSEIR_heatmaps_r} displays heatmaps of $I_\text{max}$ and $t(I_\text{max})$ as functions of $r$ and $\gamma$.
\FloatBarrier
This time, a region of the heatmap corresponding to low $r$ and high $\gamma$ has $t(I_\text{max})=0$. Similar to the previous section for regular networks, it appears that this phenomenon is related to the $R_0$ of the model. Based on the calculations in Ref.~\cite{R0_heterogeneo}, we get the following approximation of $R_0$:
\begin{equation}
R_0 = r\ T_\text{inf}\ \frac{\langle k^2 \rangle}{\langle k\rangle^2}\ \left(\langle k \rangle -1\right)
\label{R0_KSEIR}
\end{equation}
To further investigate this relationship, Figure~\ref{R0_curves} illustrates the curve of $\gamma^\text{crit}(r)$ that delineates the region with $t(I_\text{max})=0$ from Figure~\ref{KSEIR_heatmaps_r_b}, as well as the curve obtained for $R_0=1$ in Eq.~(\ref{R0_KSEIR}). Remarkably, these curves are compatible with each other, especially for $\gamma\leq 3$.

In the main article, we set the initial parameter of the degree distribution to $\gamma_0=2.5$ and the mean infection rate to $r=0.2$, which according to Eq.~(\ref{R0_KSEIR}) results in $R_0=8.74$. With these values, the $I_\text{max}$ attained is close to $0.15$ as seen in Figure~\ref{KSEIR_heatmaps_r_a}. Analogously to the previous section, we set $\tau=0.05$ so that it is approximately $I_\text{max}/3$.

\bibliographystyle{elsarticle-num}